\documentclass{aa} 
\usepackage{graphics} 
\usepackage{epsfig} 
\usepackage{amssymb}

\def\et{et al.}   
\def\etal{et al. }  
  
\def\msun{{${\rm M}_\odot$} }

\def\Z0{{$\rm\,Z_\odot$}}

\def\n#1{{~}}
\def\spose#1{\hbox to 0pt{#1\hss}} 
\def\lta{\mathrel{\spose{\lower 3pt\hbox{$\mathchar"218$}} 
     \raise 2.0pt\hbox{$\mathchar"13C$}}} 
\def\gta{\mathrel{\spose{\lower 3pt\hbox{$\mathchar"218$}} 
     \raise 2.0pt\hbox{$\mathchar"13E$}}} 
\def\h0{} \def\q0{{q$\_0$}}

\begin{document}

\title{Chemically consistent evolution of galaxies:} 
\subtitle{II. Spectrophotometric evolution from zero to high redshift}
 
\author{Jens Bicker$^1$, Uta Fritze -- v. Alvensleben$^1$, Claudia S. M\"oller$^2$, Klaus J. Fricke$^1$} 
 
\institute{$^1$Universit\"atssternwarte, Geismarlandstra\ss e 11, D--37083 G\"ottingen, Germany\\
	   $^2$Max Planck Institute for Astrophysics, Karl-Schwarzschild-Str. 1, 85741 Garching, Germany}

\offprints{J. Bicker, Universit\"atssternwarte, Geismarlandstr. 11,  
        37083 G\"ottingen, Germany   (jbicker@uni-sw.gwdg.de)} 
  
\date{Received ... , 2003; accepted ... , 2003} 
 
\authorrunning{J. Bicker et al.} 
\titlerunning{Spectrophotometric evolution with redshift} 
 
\abstract{  
The composite stellar populations of galaxies comprise stars of a
wide   range of metallicities. Subsolar metallicities become increasingly
important,  both in the local universe when going from early towards later galaxy
types   as well as for dwarf galaxies and for all types of galaxies towards higher
redshifts.  

We present a new generation of chemically consistent evolutionary synthesis   models
for galaxies of various spectral types from E through Sd. The models follow the
chemical enrichment of the ISM and take into account the increasing initial 
metallicity of successive stellar generations using recently published  metallicity
dependent stellar evolutionary isochrones, spectra and yields. 

Our first set of closed-box 1-zone models does not include any spatial   resolution or
dynamics. For a Salpeter initial mass function (IMF) the star formation rate
(SFR) and its time evolution are shown to successfully parameterise spectral
galaxy types E, ..., Sd.   We show how the stellar metallicity distribution in various
galaxy types build up with time to yield after $\sim 12$ Gyr agreement with stellar
metallicity distributions observed in our and other local galaxies.

The models give integrated galaxy spectra over a wide wavelength 
range (90.9\AA - 160$\mu$m), which for ages of $\sim 12$ Gyr are in good  
agreement not only with observed broad band colours but also with template  
spectra for the respective galaxy types.  
 
Using filter functions for Johnson-Cousins U,B,V,$\rm{R_C}$,$\rm{I_C}$, as well as
for  HST broad band filters in the optical and Bessel \& Brett's NIR  J,H,K filter
system, we calculate the luminosity and colour evolution   of model galaxies over a
Hubble time. 

Including a standard cosmological model (${\rm H_0 = 65, ~\Omega_0 = 0.1}$) and the
attenuation by intergalactic   hydrogen we present evolutionary and cosmological
corrections as well as   apparent luminosities in various filters over   the redshift
range from z $\sim 5$ to the present for our   galaxy types and compare to earlier
models using single (=solar) metallicity input physics only. We also resent a first
comparison of our cc models to HDF data. A more detailed comparison with Hubble
Deep Field (HDF) and other deep field data and an analysis and interpretation of high
redshift galaxies in terms of ages, metallicities , star formation histories and,
galaxy types will be the subject of a forthcoming paper.  

\keywords{Galaxies: evolution -- Galaxies: stellar content -- Galaxies: general -- 
Galaxies: photometry -- Galaxies: spectra -- Galaxies: redshifts --Cosmology:
observations}}

%______________________________________________________________  

\maketitle 

\section{Introduction} 
 
The number of high and very high redshift galaxies is increasing rapidly these days.
Thousands of U- and B-dropout galaxies with photometric redshift estimates are known
from the HST Hubble Deep Fields - North  and South (Williams et al. 1996, 1998, Gardner
et al. 2000, Ferguson et al. 2000)  -- and from ground-based deep surveys like the VLT-
Fors Deep Field (FDF) and others. Hundreds of these have spectroscopically confirmed
redshifts up to ${\rm z \sim 6}$ (e.g. Hu et al. 1999). Deep surveys are being
conducted at all wavelengths from UV through IR and far into the sub-mm range. These
brilliant data require mature galaxy evolution models for adequate interpretation.
Ideally, these models should cover all the observational wavelength baseline to allow
for a consistent interpretation of all the available data, be as comprehensive and
realistic as possible, and extend from very early phases at very high redshift towards
the global properties of nearby galaxies of various spectral types. Moreover, an ideal
galaxy model should be as simple as possible, involving the smallest possible number of
free parameters. A comprehensive galaxy model should describe the evolution of as many
observable quantities as possible (spectrum, luminosities, colours, emission and
absorption features for the stellar population, the gas content and a large number of
element abundances for the interstellar medium (ISM)). A realistic galaxy
evolution model should consistently take into account both the age and metallicity
distributions of the stellar populations that naturally result from any extended star
formation history (SFH). 

This is what we attempt with our chemically consistent spectrophotometric,
chemical and cosmological evolutionary synthesis model. The chemical
evolution   aspects of this model were presented  by Lindner \et\ (1999),
based on earlier less complete stellar  evolutionary tracks, in comparison
with and interpretation of the observed redshift evolution of   Damped
Ly$\alpha$ Absorber (DLA) abundances. Here, we present the spectral and
spectrophotometric aspects of our unified chemical and spectral evolutionary
synthesis code including current stellar evolutionary isochrones. 

Spectrophotometric and cosmological evolutionary synthesis models generally applied in
current interpretations of high redshift galaxy data are using solar metallicity input
physics only together with specific parametrisations for the star formation (SF)
histories of various spectral types (e.g. Bruzual \& Charlot 1993, Bressan et al. 1994,
Guiderdoni \& Rocca -- Volmerange 1987, 1988, Fioc \& Rocca -- Volmerange 1997,
Poggianti 1997). Broad stellar metallicity distributions extending down to fairly low
[Fe/H] are observed in local galaxies of various types. We will show in how far the
consideration of a realistic metallicity distribution among the stars will affect model
predictions for high redshift galaxies.  

The first attempts to account for non-solar abundances and their impact on the  
photometric evolution of galaxies go back to Arimoto \& Yoshii (1986).   Einsel
\et\ (1995) used more recent and complete stellar evolutionary   tracks and colour
calibrations for initial stellar metallicities   $10^{-4}$ to $4 \cdot 10^{-2}$ to
describe in a chemically consistent way the photometric evolution of galaxy   types
E through Sd. In M\"oller \et\ (1997) we introduced the   concept of   chemical
consistency into the spectrophotometric evolution of galaxies   using Kurucz's
(1992) model atmosphere spectra for a range of stellar   metallicities and
investigated the time evolution of ISM metallicity and   luminosity-weighted mean
stellar metallicities in various wavelength   bands. For models that well agree
with observed template spectra   (Kennicutt 1992) of various types (E, Sb, Sd) we
gave decompositions of   the total light emitted at wavelengths from U through K in
terms of   luminosity contributions from various metallicity subpopulations.   This
clearly showed the considerable widths of the metallicity   distributions in all 3
galaxy types as well as the importance of luminosity  contributions from stellar
populations with subsolar metallicities.  
Recently V\'azquez et al. (2003) presented their evolution synthesis code SPECTRAL
in  application to the irregular galaxy NGC 1560 that also accounts for the the
simultaneous presence of stars of different metallicities. They use stellar 
evolutionary input of the Geneva group and describe the spectral evolution of NGC 1560
on the basis on chemical evolution models by  Carigi et al. (1999).

This paper is organised as follows.   In Sections 2 and 3 we present our chemically
consistent galaxy evolution model   and the various pieces of input physics for
different metallicities   that it uses.   Sect. 4 gives a comparison of our model
colours and   spectra with observations of nearby galaxies to show that   -- after
a Hubble time -- our models do reproduce the observed   properties of typical
local galaxy types E ... Sd. In Sect. 5 we   present the results for our models E,
Sb, Sd in terms of spectra at   various ages, apparent magnitudes, evolutionary
and cosmological   corrections (including attenuation) in wavelength bands ${\rm
UBVR_cI_cJHK}$ and   a series of HST broad band filters as a function of
redshift.   Results are presented for the redshift range   ${\rm 0 \leq z \leq
4.8}$ for a cosmological model  ${\rm H_0 = 65, ~\Omega_0 = 0.1}$, with  a
redshift  of 5 assumed for galaxy formation in both cases.  

We compare our chemically consistent models to models using solar   metallicity
input physics only and present a first comparison   with HDF galaxies with
spectroscopic redshifts in Sect. 6.  Conclusions are given in Sect. 7.  A further
set of models which also includes the influence of dust in a   chemically
consistent way will be presented a separate paper (see M\"oller \etal 2001a,b for
first results).   A detailed interpretation of the HDF data with   the models
presented here will be the subject of a forthcoming paper.  

\section{Metallicity observations in galaxies} 
\label{observations} 
 
For quite some time, observational evidence has been   accumulating for sometimes very
broad metallicity distributions of   stars in normal galaxies.  Stars in the Milky Way
disk and halo span a range of at least 4 orders of magnitude in metallicity $-4.2 \le
{\rm [Fe/H]} \le +0.3$. While some years ago, the focus was on super-solar  
metallicities e.g. in (the centres of) massive ellipticals, bulges, X-ray halos  
around ellipticals, and the hot intra cluster medium (ICM), by today, it is clear that the
{\bf average   metallicities} in all those cases are {\bf subsolar}. The sun, our
reference star, stands out in metallicity among solar   neighbourhood stars. For F, G,
K dwarfs the [Fe/H] distributions extend from ${\rm -0.8~ to~ +0.4}$ (Rocha-Pinto \&
Maciel 1998), while B-stars show ${\rm \langle [O/H] \rangle = -0.31}$
(Kilian-Montenbruck \etal 1994). When averaged over ${\rm
1~R_e,~with~R_e~:=~effective~radius}$, line strength   gradients in ellipticals
indicate ${\rm \langle Z_{\ast} \rangle \sim (0.5 - 1)\cdot Z_{\odot}}$   (Carollo \&
Danziger 1994). For stars in the bulge of our Milky Way   ${\rm \langle Z_{\ast}
\rangle \sim (0.3 - 0.7)\cdot Z_{\odot}}$ (e.g.   McWilliam \& Rich 1994, Sadler \etal
1996, Ramirez \etal 2000),   for the X-ray gas halos around elliptical galaxies ASCA
observations give   ${\rm 0.1 \leq [Fe/H] \leq 0.7}$ (e.g. Loewenstein 1999).  

Characteristic HII region abundances (i.e. measured at ${\rm 1~R_e}$), which   give
an upper limit to the average gas phase abundance, range from ${\rm Z \gta
Z_{\odot}}$   for Sa spirals down to ${\rm < \frac{1}{2}Z_{\odot}}$   for Sd
galaxies (e.g. Oey \& Kennicutt 1993, Zaritsky \etal 1994, Ferguson \etal 1998,
van   Zee \etal 1998). Locally, dwarf irregular galaxies have metallicities
in   the range (2 -- 30)\% ${\rm Z_{\odot}}$ (e.g. Richer \& McCall 1995). The
first spectra of   Lyman break galaxies at redshifts ${\rm z \sim 3 - 4}$ have
shown that their metallicities, derived   from stellar wind features, are
considerably subsolar, sometimes even sub-SMC (Lowenthal   \etal 1997, Trager \etal
1997, Pettini \etal 2000, Teplitz \etal 2000).   Neutral gas in damped Ly$\alpha$
absorbers observed to ${\rm z > 4}$ shows   abundances in the range ${\rm -3 \lta
[Zn/H] \lta 0}$ (e.g Pettini \etal 1997, 1999, Lindner \etal 1999).

%______________________________________________________________ 
\section{Chemically consistent galaxy evolution models}  
\label{model_description} 

From a very principle point of view it is clear that in contrast to single   burst
stellar populations like star clusters, any galaxy with a   star formation history
extending over one to several Gyr, i.e.  much longer than the   lifetime of massive
stars,  will have a stellar   population that is composite not only in   age but
also in metallicity -- as confirmed by the observations cited above.   This is what
we intend to account for in our chemically   consistent galaxy evolution models.  
The basic concept of our evolutionary synthesis model for galaxies has been
described in detail by  Fritze -- v. Alvensleben \& Gerhard (1994),  the extended
version allowing for a {\bf c}hemically {\bf c}onsistent (={\bf cc})   modelling is
described by Einsel et al. (1995) for the   photometric evolution and in detail
by   M\"oller \et\ (1997),   M\"oller \et\ (1999a),   Fritze -- v. Alvensleben \et\
(1999) for the spectral and   spectrophotometric evolution and by Lindner \et\
(1999) for the chemical evolution.

In the following we briefly outline   the principle of the new concept of chemical
consistency which we   consider an important step towards a more realistic galaxy
modelling.   In contrast to single burst single metallicity stellar populations   like
star clusters (Schulz \etal 2002)   our chemically consistent galaxy evolution model,  
solving a modified set of Tinsley's equations with metallicity dependent   stellar
yields, follows the   metal enrichment of the ISM and accounts for the increasing  
initial metallicity   of successive stellar generations, both with respect to   the
evolution of ISM abundances (Lindner \et\ 1999) and   to the spectral evolution as
presented here.  The evolution of each star  is followed in the HR diagram from birth
to its final phases  according to stellar evolutionary tracks appropriate for its  
initial metallicity  so that at each timestep the distribution of all stars over the 
HRD is known.  The evolution of the HRD   population is followed with  various sets of
stellar isochrones from the Padova group   for five different metallicities from  $Z= 4
\cdot 10^{-4}$ to $5 \cdot 10^{-2}$. Stellar subpopulations formed with initial
metallicities in between two of the 5 discrete metallicities of the Padova isochrones
are described by two components from the two adjacent metallicities. The relative
contributions of these two components are weighted by the inverse of the logarithmic
differences between the metallicity of the subpopulation and the metallicities of the
adjacent isochrones.  At any timestep the HRD population is used to synthesise an
integrated   galaxy spectrum from a library of stellar spectra. This library  
comprises   stellar model atmosphere spectra  from UV to the IR for all spectral types
and luminosity classes for   5 metallicities (Lejeune \et\ 1997, 1998).   The total
galaxy spectrum is obtained by summing the isochrone spectra,   weighted by the star
formation rate at birth of the stars on the respective isochrone  for each metallicity
and,   finally, by co-adding the spectra of the   various single metallicity
subpopulations.  Combining the spectrophotometric time evolution with a cosmological  
model and some assumed redshift of galaxy formation we calculate the   evolutionary and
cosmological corrections as well as the evolution of   apparent magnitudes from optical
to NIR for various galaxy types taking into account the attenuation of the emitted
galaxy light by intervening HI (cf. Sect. 3.4).

%- - - - - - - - - -  - - - - - - - - - - - - - - - - - - - - - - - - - - - 
\subsection{Input physics} 
\label{input_physics} 
 
In an attempt to keep the number of free parameters as small as possible,   our
models are calculated as closed boxes with instantaneous and perfect   mixing of
the gas.  We use isochrones from stellar evolutionary tracks provided by 
Bertelli \etal (1994) (Padova group) in the version from November 1999 that
include the thermal pulsing AGB phase as described in Schulz \etal (2002). The
stellar lifetimes for the various metallicities are fully taken into account
in   our description of the ISM enrichment.   

The only stellar library covering the whole range of stellar metallicities,  
spectral types, and luminosity classes   is the library of model atmospheres by
Lejeune \etal (1997, 1998), based on the   original library of Kurucz (1992).  
Its wide wavelength range from 0.09 to 160000 nm extending far into the UV
allows us to calculate   cosmological corrections even in   the U band out to
very high redshift z$\sim $5.  For stars hotter than 50.000 K, the highest
effective temperature of Lejeune \et's library, we use black body spectra.
To follow in detail the chemical evolution of the ISM, Lindner et al. (1999)
included   stellar yields from  Woosley \& Weaver (1995) and van den Hoek \&
Groenewegen (1997)  for various metallicities. For the models in this
paper we are not interested in the detailed chemical evolution, We only use the 
time evolution of the global metallicity Z to know when to switch from one isochrone
to the next (= more metal rich).

With this new set of metallicity dependent input physics our models are now 
chemically consistent both with respect to the spectrophotometric and to the
chemical   evolution.  

The various spectral galaxy types of the Hubble Sequence of normal galaxies  
are described by their respective appropriate galaxy-averaged  star formation
histories. For spheroidal galaxies (E) models use a {\bf S}tar {\bf F}ormation
{\bf R}ate   SFR ${\rm \sim e^{-\frac{t}{t_{\ast}}}}$ with an e-folding time  
${\rm t_{\ast} = ~1}$ Gyr.   Following Kennicutt (1998) we assume for the
spirals a  SFR linearly proportional to the gas-to-total mass ratio with  
characteristic timescales for the transformation of gas into stars ranging from
${\rm t_{\ast} = ~4}$ Gyr for Sa through ${\rm t_{\ast} > ~15}$ Gyr for Sd
spectral types.  

We use a standard Salpeter IMF from lower to upper mass limits ${\rm m_l=0.08}$\msun
to ${\rm m_u\sim 70~M_\odot}$, as given by the isochrones. The IMF is normalised to a
fraction of  visible mass of ${\rm FVM= 0.5}$ to match the mass-to-light (${\rm M/L}$)
ratios for today's ($\sim 12$ Gyr old) galaxies.

A Scalo (1986) or a Kroupa (1993) IMF would produce a lower number of massive stars than the
Salpeter IMF. Lindner \etal (1999) have shown that the metallicity evolution of spirals
is too slow for a Scalo IMF as compared to observations. As compared to the Salpeter IMF
that we use, these other IMFs, whithout adjustment of the SFRs, would produce a larger fraction of 
low metallicity stars and, hence, lead to slightly bluer galaxy colours.

%-  -  -  -  -  -  -  -  -  -  -  -  -  -  -  -  -  -  -  -  -  - 
\subsection{Filters and calibrations} 

We calculate the apparent magnitudes ${\rm m_{\lambda}}$, evolutionary and
cosmological - corrections  for the UBV- Johnson \& ${\rm R_CI_C}$ Cousins filter
system which is taken from Lamla (1982), for the JHK filters given by Bessel \&
Brett (1988), and for all broad band HST WFPC2 filters. Magnitudes in all filters
are calibrated in the VEGAMAG system. On the basis of the time
evolution of the model spectra we provide the evolution in other filter systems,
easily calculated by directly folding the filter and detector response curves with
the model spectra.   

%-  -  -  -  -  -  -  -  -  -  -  -  -  -  -  -  -  -  -  -  -  - 

\subsection{Cosmological model} 
 
In order to compare our models with data of high redshift galaxies, 
we transform the spectrophotometric time evolution into a redshift  
evolution with a set of cosmological  
parameters (${\rm H_0,\Omega_0, z_f}$),  where ${\rm z_f}$ is the  
redshift of galaxy formation. 
 
The age of a galaxy at redshift ${\rm z = 0}$ is given by 
 
\begin{center}  ${\rm t_0 := t_{gal}(z=0) := t_{Hubble}(z=0) - t_{Hubble}(z_f)}$ \end{center}

After ${\rm t_{gal}(z=0)}$ models are normalised  
to the observed average absolute luminosities ${\rm M_B^{\ast}}$ of the respective galaxy types  
in Virgo (cf. Sandage et al. 1985a,b). 
To obtain the observable apparent magnitudes ${\rm m_{\lambda}}$  
from the absolute magnitudes ${\rm M_{\lambda}}$ given by our models,  
we calculate the evolutionary (${\rm e_{\lambda}}$) 
and cosmological corrections (${\rm k_{\lambda}}$) and the bolometric distance modulus 
BDM(${\rm H_0,\Omega_0}$):  
 
\begin{center} ${\rm m_{\lambda}(z) = M_{\lambda}(z=0,~t_0) + BDM(z) + e_{\lambda}(z) +  
                 k_{\lambda}(z) }$ \end{center} 
 
The cosmological or k - correction ${\rm k_{\lambda}}$ in any wavelength band $\lambda$  
describes the effect of the expanding universe, which redshifts a  
galaxy spectrum of local age ${\rm t_0}$ to some redshift z 
 
\begin{center} ${\rm k_{\lambda}(z) := M_{\lambda}(z,~t_0) - M_{\lambda}(0,~t_0)}.$ \end{center} 
 
${\rm k_{\lambda}}$ can also be calculated from observed galaxy spectra.  
In this case, the maximum redshift to which 
 this is possible depends on how far into the UV the observed spectrum extends. Our model galaxy  
 spectra at ${\rm z=0}$ extend from 90 \AA \ through  160 $\mu$m and, hence, allow for  
cosmological corrections in optical bands up to ${\rm z \gg 10}$.  
 
The difference in absolute luminosities between two galaxies at the same redshift but 
with different ages is described by the evolutionary correction  
 
\begin{center} ${\rm e_{\lambda}(z) := M_{\lambda}(z,t_{gal}(z)) - M_{\lambda}(z,~t_0)}.$ \end{center} 
 
Evolutionary corrections, of course, cannot be given without an evolutionary
synthesis model.   It will be shown in Sect. 5 for which galaxy types at which
redshifts evolutionary   corrections become important.  

It is important to stress that both the evolutionary and the cosmological
corrections do not only depend on the cosmological parameters but also on the
SFH, i.e. on the spectral type of the galaxy.  

In this paper we present cosmological and evolutionary corrections and apparent
magnitudes for cosmological parameters ${\rm (H_0,~\Omega_0) ~=~
(65,~0.1)}$ and the formation of galaxies at redshift ${\rm z_f = 5}$. 
This cosmology gives a galaxy age of about 12 Gyr after a
Hubble time   and, hence, makes sure that the agreement with colours and spectra
of nearby galaxies at z=0 is given. Any
combination of H$_0$ and $\Omega_0$ which gives a local galaxy age   ${\rm t_{gal}
\lta 10}$ Gyr can   be excluded because the very red colours of ellipticals can
only be reached after $\gta 12$ Gyr   and globular cluster ages also are of
order 12 to 15 Gyr. In particular, high values of   ${\rm H_0 > 80}$, even in
combination with low values for ${\Omega_0}$, yield galaxy ages less  than 10
Gyr.

%-  -  -  -  -  -  -  -  -  -  -  -  -  -  -  -  -  -  -  -  -  - 
\subsection{Attenuation}  

For very distant galaxies the cumulative effect of neutral hydrogen
stochastically   distributed along our lines of sight, mostly in the form of
Ly$\alpha$ clouds,   significantly attenuates the emitted light at wavelengths
shorter than rest-frame   Ly$\alpha$ at 1216 \AA. From his analysis of a
large number of lines of sight   to distant and very distant quasars, Madau
(1995) derived a statistically averaged   attenuation correction of the form
$att(\lambda,~z)$ which we include into the   cosmological corrections of our
models. The effect of attenuation becomes visible in   U at ${\rm z \gta 2}$, in
B at ${\rm z \gta 2.5}$, in V at ${\rm z \gta 3.5}$, etc.

%-  -  -  -  -  -  -  -  -  -  -  -  -  -  -  -  -  -  -  -  -  - 
\subsection{Discussion of models}\label{discuss_model} 
 
Our models describe the global spectrophotometric evolution of field  galaxies.
In particular, our E model re\-presents the classical model for a normal
elliptical  galaxy with average luminosity and metallicity. A hierarchical or
major  merger origin of ellipticals is not investigated here.
%****
The closed-boxes we assume for our spiral models clearly  are a poor approximation
only motivated by our intend to keep this first models as simple as possible an the
number of free parameters to a minimum. For a hierarchically accreting spiral the
total SFH of its ensemble of subclumps on reasonable time averages should not
differ mutch from what our simplified models assume in order to get the correct
colours and spectra at ${\rm t_0}$. The same had been to be true for the chemical
evolution of an ensemble of subclumps as compared to our simplified closed-box
models by Lindner \et\ (1991). The colour evolution hence is not significant
affected by our simplification and, as far back as the mass of a realistic
accreting galaxy is of order $\ge 30$\% of its mass today the differences in
apparent luminosities can be estimated to be smaller than differences between
different galaxy types.
%****
In a previous paper we have  shown that our models well describe the stellar
metallicity as observed by means of    absorption indices (M\"oller et al. 1997)
and the chemical evolution  of nearby and high redshift spiral galaxies (Lindner et
al. 1999). 

In this paper we restrict ourselves to models without dust. The effect of 
dust absorption is analysed in M\"oller et al. (1999b) with our old track
based models. A set of  models including dust will be presented by M\"oller
et al. (in prep.).

%------------------------------------------------------------------------- 
\section{Comparison with nearby galaxies}\label{nearby_gals} 
\subsection{Colours} 
 
Star formation histories for our model galaxies are chosen such that after a Hubble
time or,   more precisely, after the evolution time of a galaxy from its  
formation at redshift ${\rm z_f}$ to the present at ${\rm z = 0}$ as given by the
cosmological model, the colours of our model galaxies agree with those observed for
nearby galaxies of the respective type by Buta et al. (1995), e.g. ${\rm (B-V) =
0.92,~0.61,~and~0.50}$ for E, Sb and Sd spectral types, respectively. As a
consequence, however, our ${\rm V-K}$ colours are then bluer by up to ~0.3 mag for
Es and by up to ~0.7 mag for late-type spirals as compared to the observations of
Aaronson (1978). This is partly due to the fact that our models describe the
integrated colour of the entire stellar population, while the observations refer to
the inner parts of the galaxies. Taking into account typical observed gradients  
in ${\rm V-K}$ (c.f. Fioc \& Rocca - Volmerange 1999) brings our ${\rm V-K}$ colours into
better agreement with observations, but there still remains a difference of about
~0.5 mag for late-type spirals. As we will show in M\"oller \etal ({\sl in prep.}),
the inclusion of reasonable amounts of dust will bring models into very good  
agreement with observations over the entire wavelength range from UV through K.  

%------------------------------------------------------------------------- 
\subsection{Spectra} 
 
\begin{figure} 
%\hspace{-0.5cm} 
\resizebox{7.0cm}{!}{\includegraphics{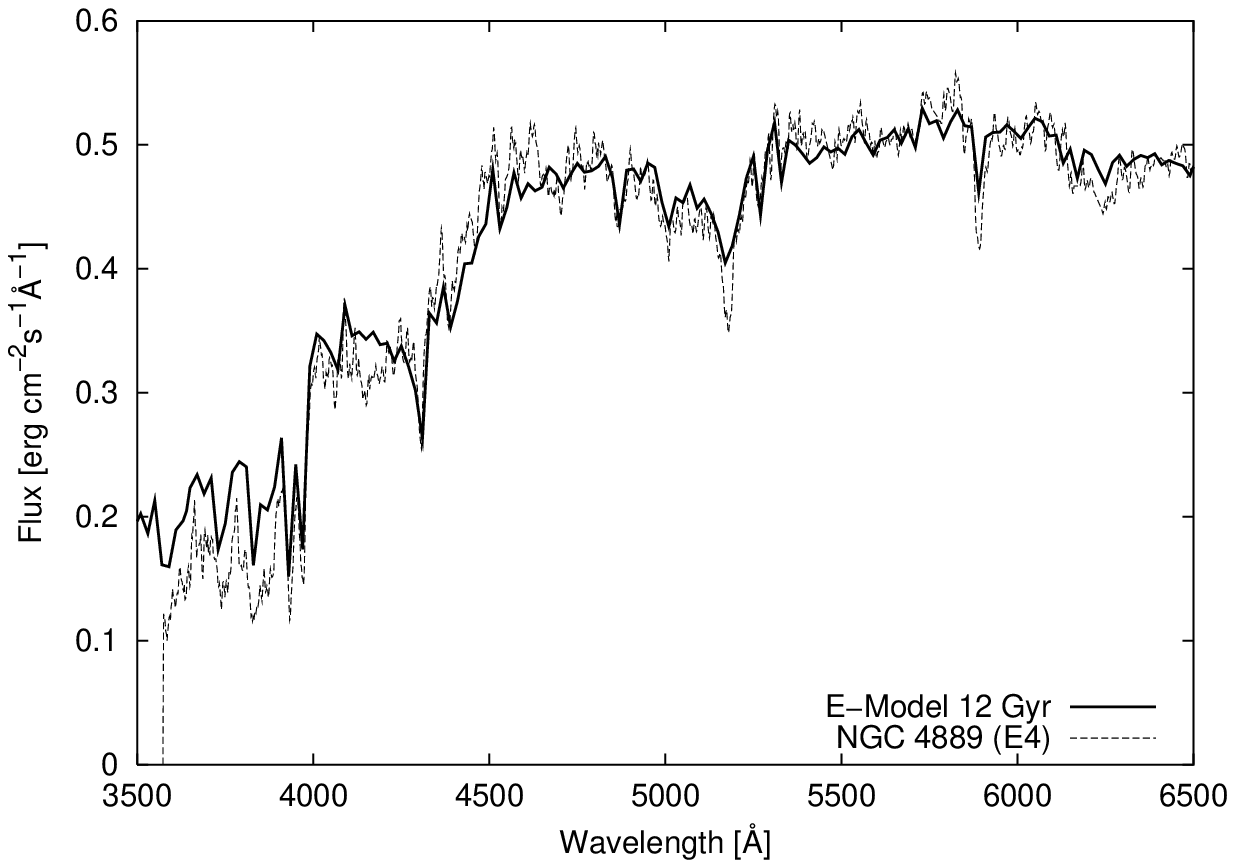}} 
\resizebox{7.0cm}{!}{\includegraphics{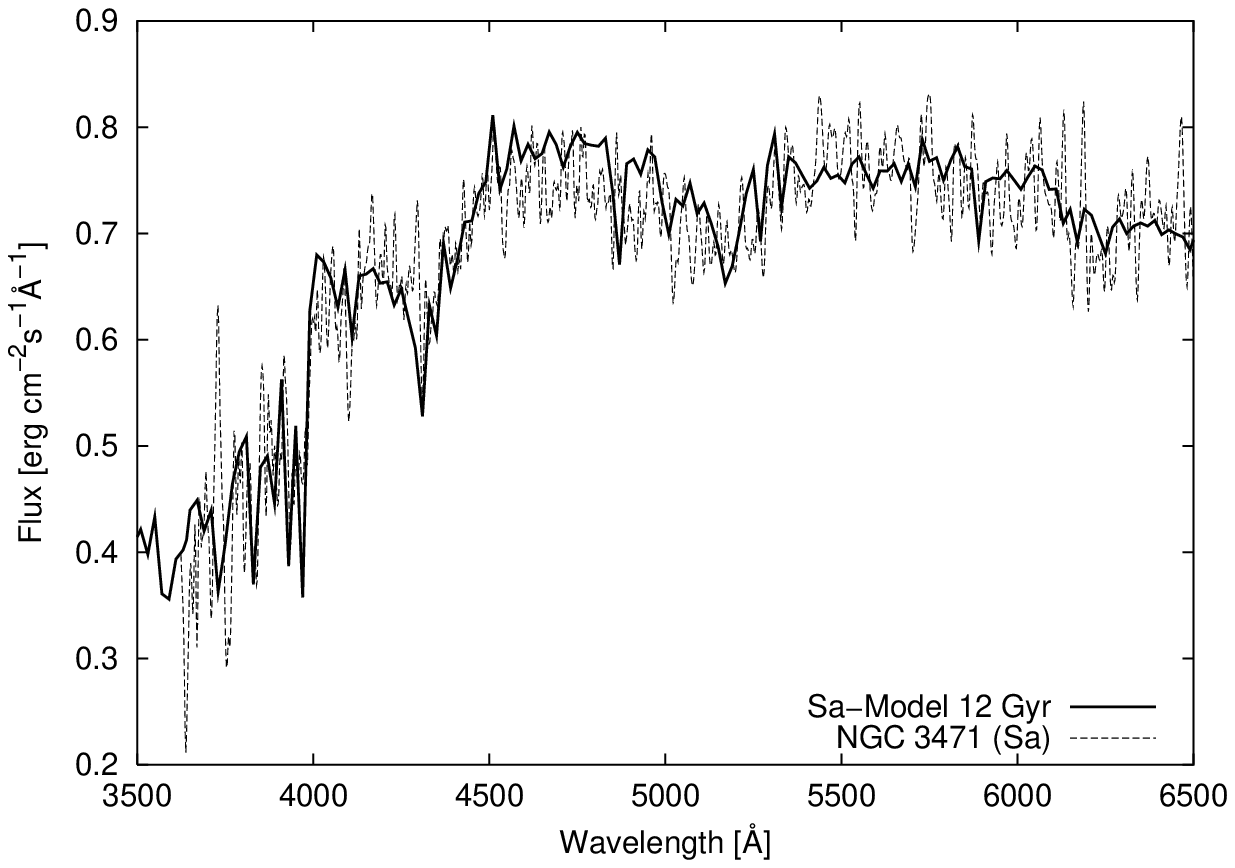}} 
\resizebox{7.0cm}{!}{\includegraphics{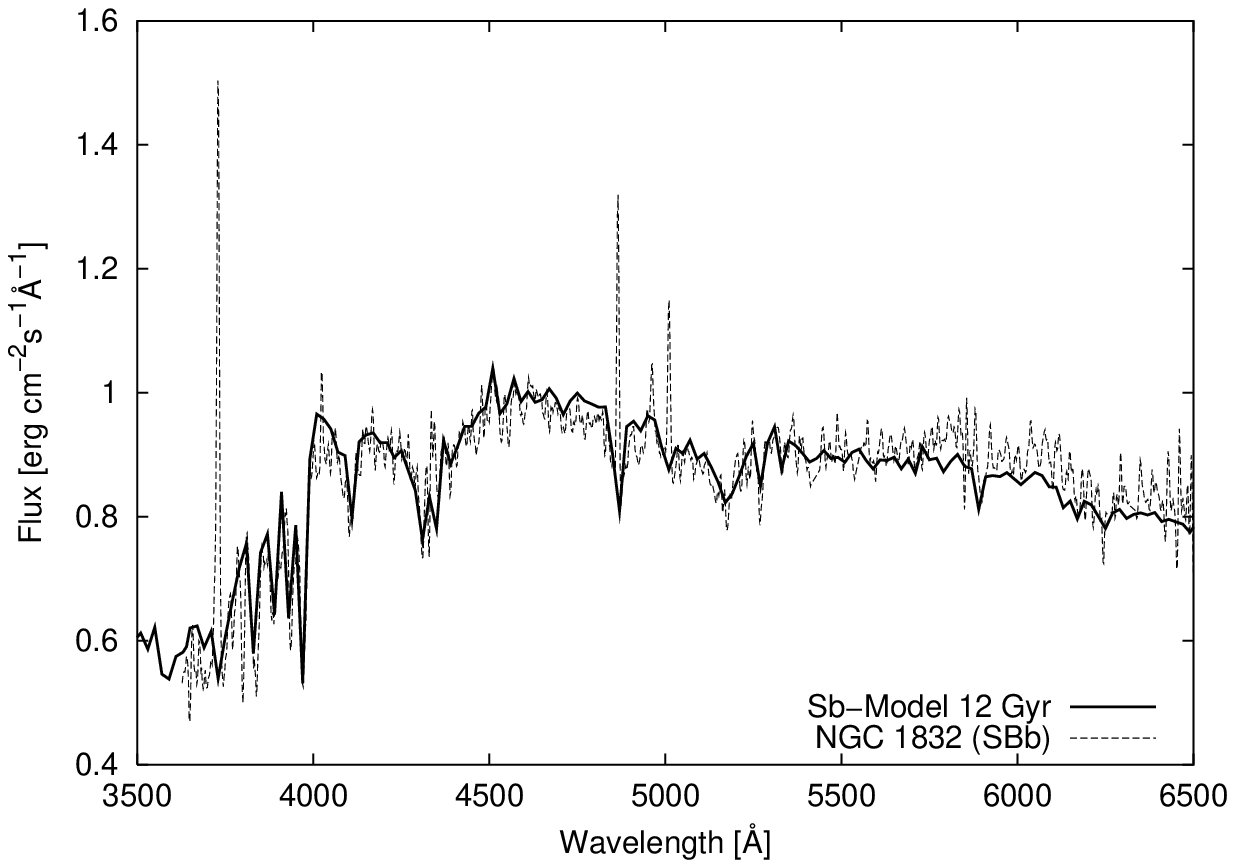}} 
\resizebox{7.0cm}{!}{\includegraphics{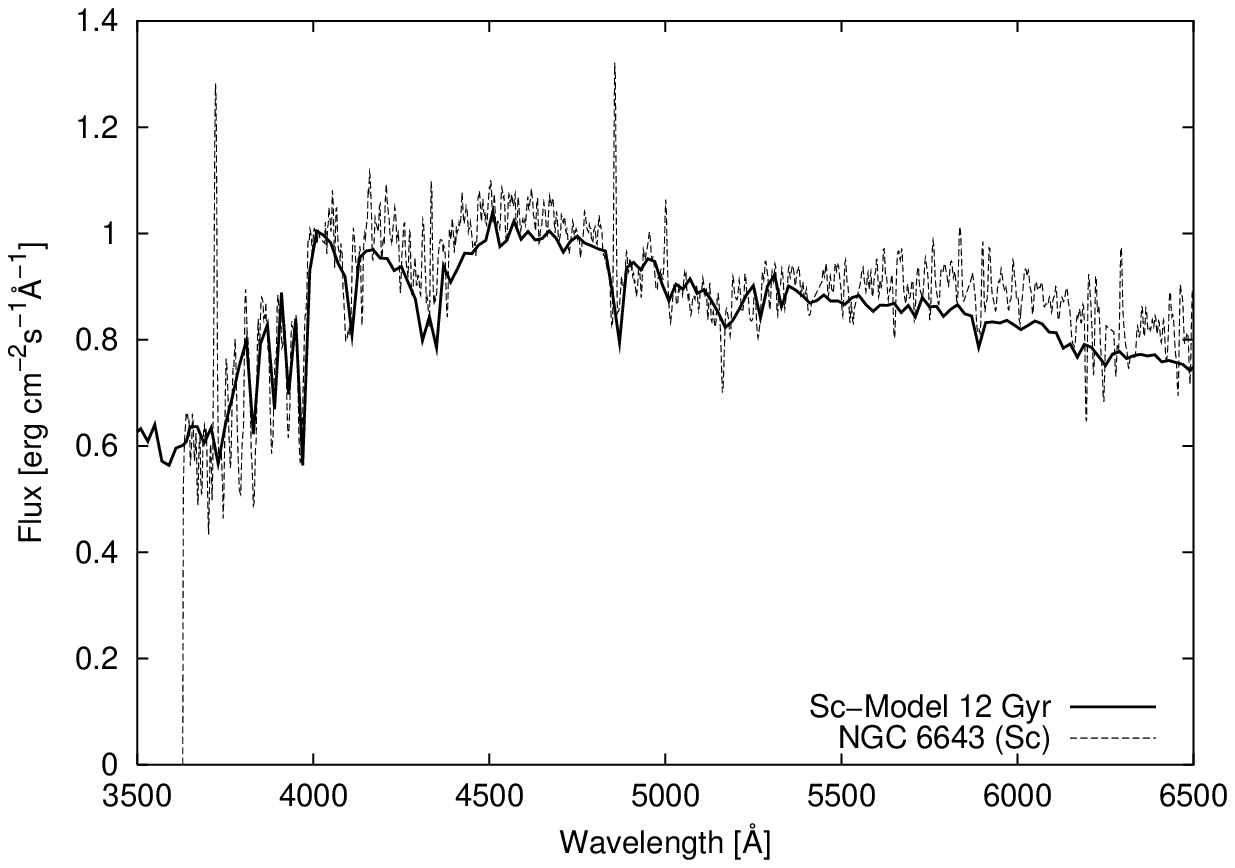}} 
\resizebox{7.0cm}{!}{\includegraphics{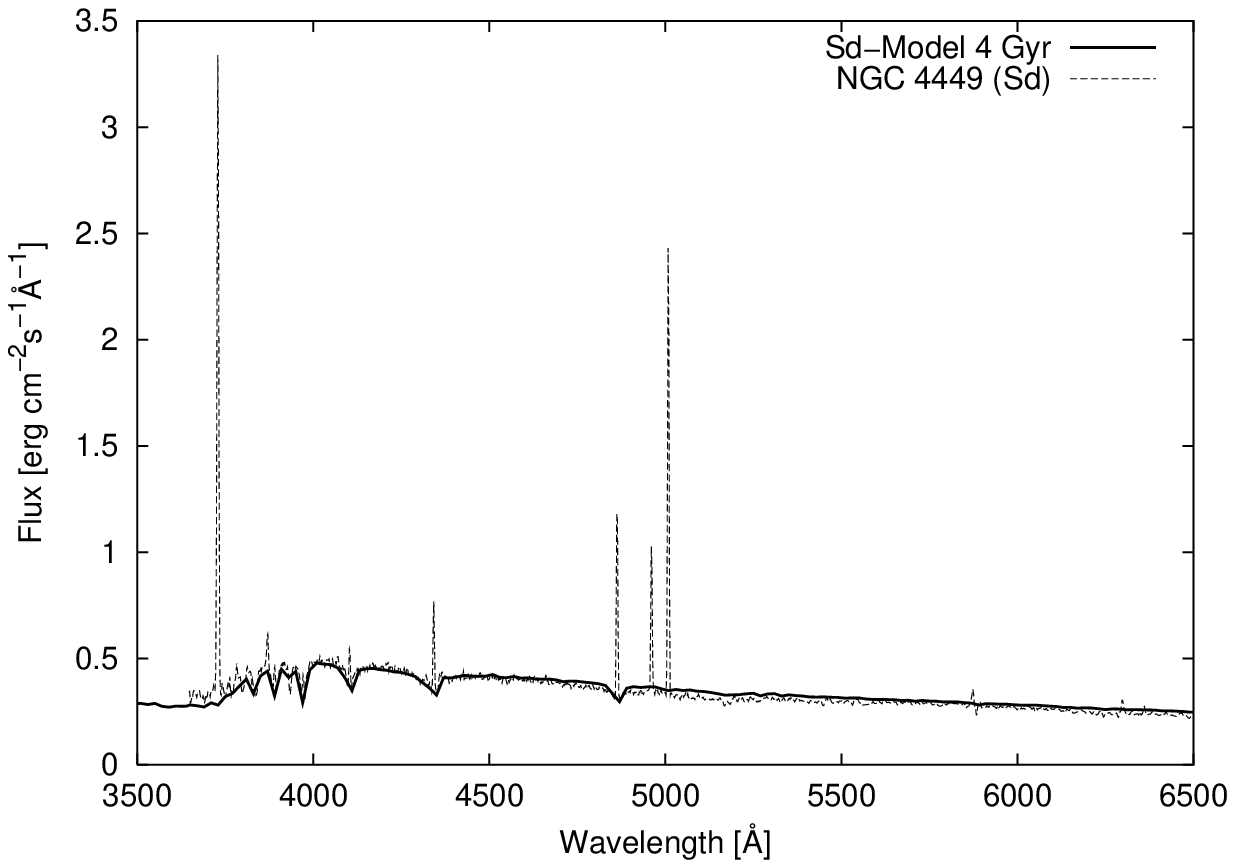}} 
\caption{Comparison of our model spectra (solid) at an age T${_{gal}}\sim 12$ Gyr (E, Sa, Sb, Sc) and T${_{gal}}\sim 4$ Gyr (Sd) with Kennicutt's template spectra (dotted).} 
\label{spectempl} 
\end{figure} 
 
In Fig. \ref{spectempl} we compare our model spectra with templates from Kennicutt's
(1992)  atlas for various spectral types. Note the very good agreement between our
models E, Sa, Sb, Sc with ages of about 12 Gyr with the observed spectra of NGC 4889
(E4), NGC 3471 (Sa), NGC 1832 (SBb), and NGC 6643 (Sc). The spectrum of NGC 4449 (Sd)
is best modelled by our Sd galaxy with an age of about 4 Gyr in agreement with results
found by Bruzual \etal (1993)
 
We like to point out that the  spectral differences among galaxies of the same
type in Kennicutt's library are larger than the differences between the template
galaxies and our models. 

Note that we have not included the gaseous emission-lines or continuum-lines in our
models yet.  In a next step, gaseous lines and continuum emission will be included in
our cc models of actively star forming galaxies on the basis of metallicity dependent
Lyman continuous fluxes and line ratios as already shown for single burst single
metallicity models by Anders \et\ (2003)  

\begin{figure} 
\hspace{-0.5cm} 
\resizebox{7.5cm}{!}{\includegraphics{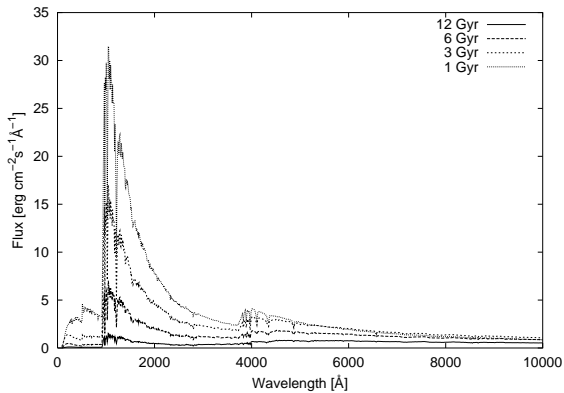}} 
\caption{Time evolution of Sa model spectra for ages 1, 3, 6 and 12 Gyr (bottom to top). Fluxes are in units 
of erg cm$^{-2}$ s$^{-1}$ \AA$^{-1}$, wavelength in \AA .} 
\label{spectimeevol} 
\end{figure}

The complete set of galaxy model spectra for the various spectral types and ages between 4 Myr to
15 Gyr are given in machine readable tables and can also be found on our homepage 
http://www.uni-sw.gwdg.de/$\sim $galev.  In Fig.\ref{spectimeevol} we show the time evolution of
model spectra,  on the example of an Sa model at 1, 3, 6, and 12 Gyr. 
 
%______________________________________________________________ 
\subsection{Metallicities} 
 
Our code follows the evolution of both the average ISM metallicity and the  
luminosity-weighted metallicities of the stellar population as seen in different
bands.  For a galaxy age of 12 Gyr, the average ISM metallicity is ${\rm \langle
Z_{ISM} \rangle = Z_{\odot}}$  for E galaxies, ${\rm \langle Z_{ISM} \rangle = 1.5
\cdot Z_{\odot}}$ for Sa, ${\rm \langle Z_{ISM} \rangle = 0.8 \cdot Z_{\odot}}$ for
Sb, ${\rm \langle Z_{ISM} \rangle = 0.5 \cdot Z_{\odot}}$ for Sc,  and   ${\rm \langle
Z_{ISM} \rangle = 0.25 \cdot Z_{\odot}}$ for Sd spirals, respectively.   These values
are in good agreement with  observations of the characteristic ($=$ measured at ${\rm
\sim 1~R_{eff}}$)   HII region abundances in the respective spiral types (e.g. Oey \&
Kennicutt 1993, Zaritsky et al. 1994,   Phillips \& Edmunds 1996, Ferguson \etal 1998,
van Zee \etal 1998). (See also Sect. 2).   

Depending on the SFH of the galaxy, the average stellar metallicity may differ by  
various degrees from the ISM metallicity and may also be different in different
wavelength   regions where stars of various masses, ages, and hence metallicities
dominate the light   (cf. M\"oller \etal 1996).  

\begin{figure} 
\resizebox{7.5cm}{!} {\includegraphics{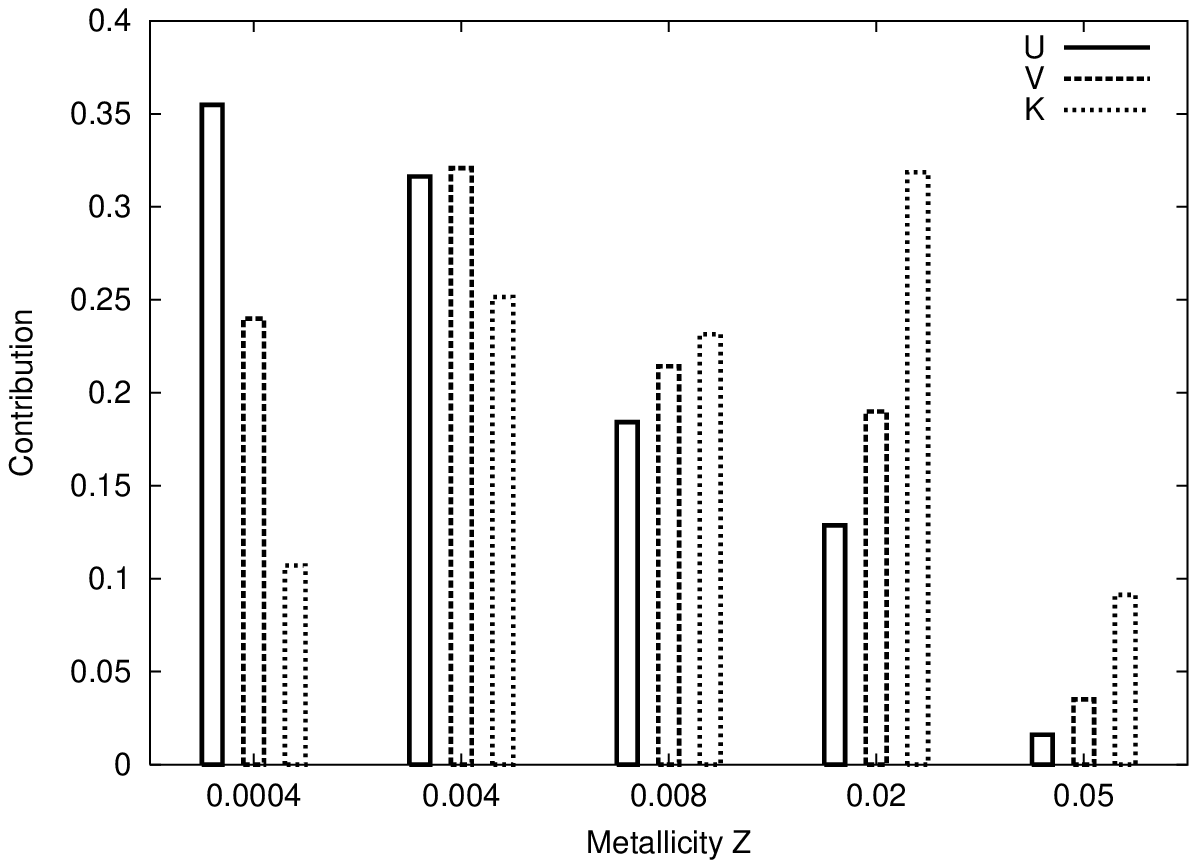}} 
\resizebox{7.5cm}{!} {\includegraphics{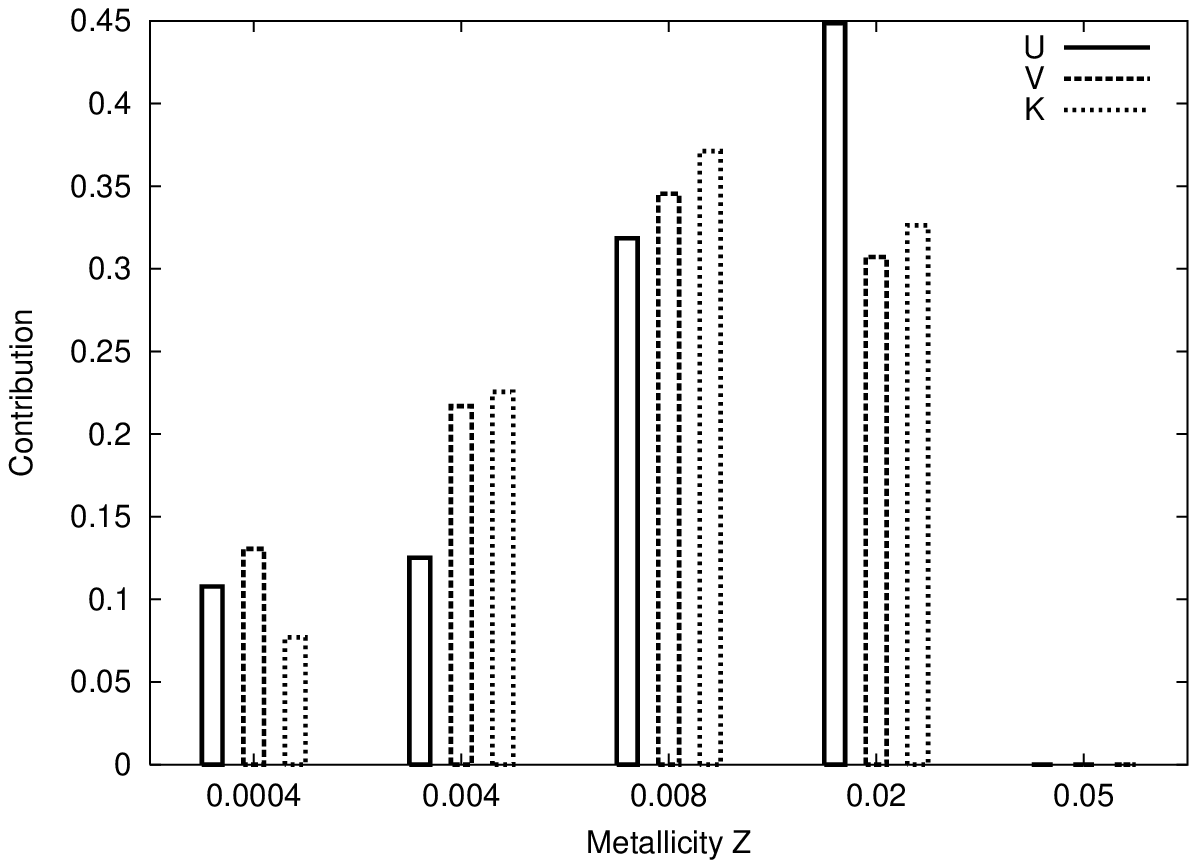}} 
\caption{Relative luminosity contributions to U, V, K bands from stellar subpopulations of various  
metallicity for 12 Gyr old E (top) and Sb (bottom) model.} 
\label{hist12} 
\end{figure}  
 
\begin{figure} 
\resizebox{7.5cm}{!} {\includegraphics{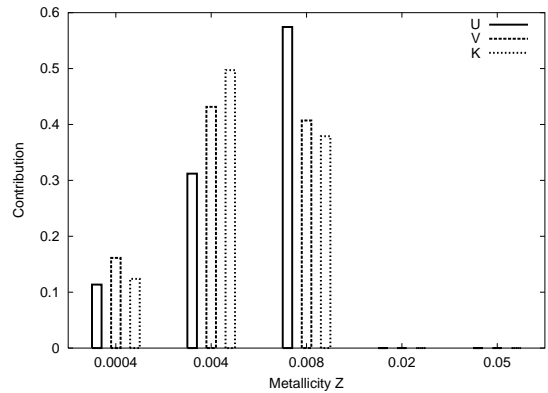}} 
\caption{Relative luminosity contributions to U, V, K bands from stellar subpopulations  of various
metallicity for a 6 Gyr old Sb model.} 
\label{hist6}
\end{figure} 
 
In Fig. \ref{hist12} we show the relative luminosity contributions to U, V, and
K bands   from stars of our 5 different  metallicity subpopulations for 12 Gyr
old E  and Sb models.   For each wavelength band, the sum of the contributions
from the different   metallicities adds up to 100 \%.   Note the broad stellar
metallicity distribution of the E model   extending from Z=$4\cdot 10^{-4}$ to
Z=0.05 in good agreement with observed stellar metallicity distributions in
resolved nearby ellipticals and bulges   (cf. McWilliam \& Rich 1994).  

The distribution differs from band to band. E.g., stars with low metallicity,  
e.g. ${\rm Z = 0.0004}$, contribute about 3 times more  light to the U- than to
the K-band, while these relative contributions are   reversed for stars of
higher metallicity.  

For models with different SFHs the stellar metallicity distributions are,   of course,
quite different. For stars in our global Sb model e.g., the metallicity   distribution
does not extend beyond ${\rm \sim Z_{\odot}}$, and the differences between the relative
contributions of a subpopulation of given metallicity to different bands are smaller
than in the E-model. For Sd models, the   stellar metallicity distribution is sharply
peaked at ${\rm Z \lta \frac{1}{2}~Z_{\odot}}$ with   small differences only between
the different wavelength bands and also to the ISM metallicity.   This is readily
understood as a consequence of the long star formation timescale in Sd galaxies   (cf.
M\"oller \etal 1997 for details).

Similar to Fig. \ref{hist12}b, Fig. \ref{hist6} shows the relative luminosity 
contribution of stellar subpopulations of different metallicities to the light  in U,
V, and K emitted by an Sb galaxy, now at a younger age of 6 Gyr only. By that age --
corresponding to a redshift ${\rm z\sim 0.5}$ in our cosmology --, no stars of solar
metallicity were present in this type of galaxy.     

\begin{figure} 
\resizebox{7.5cm}{!} {\includegraphics{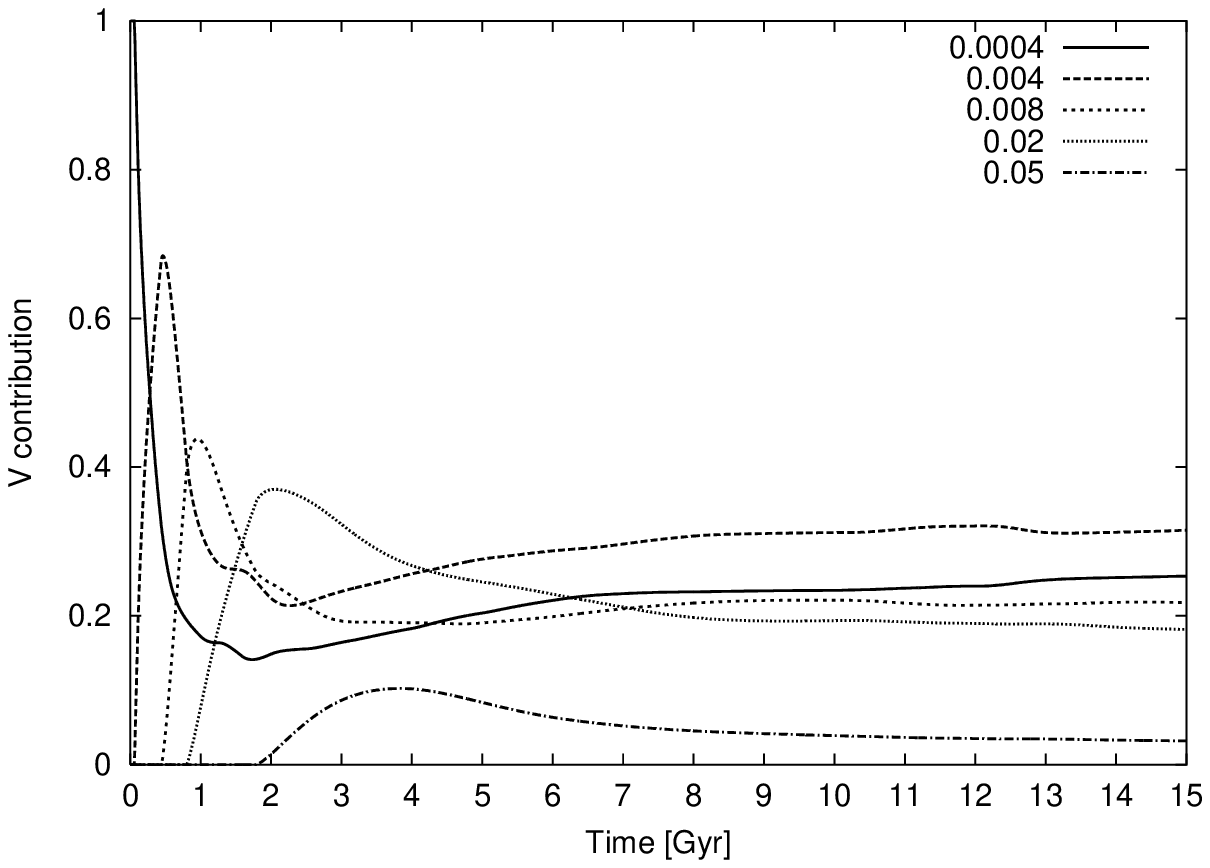}} 
\resizebox{7.5cm}{!} {\includegraphics{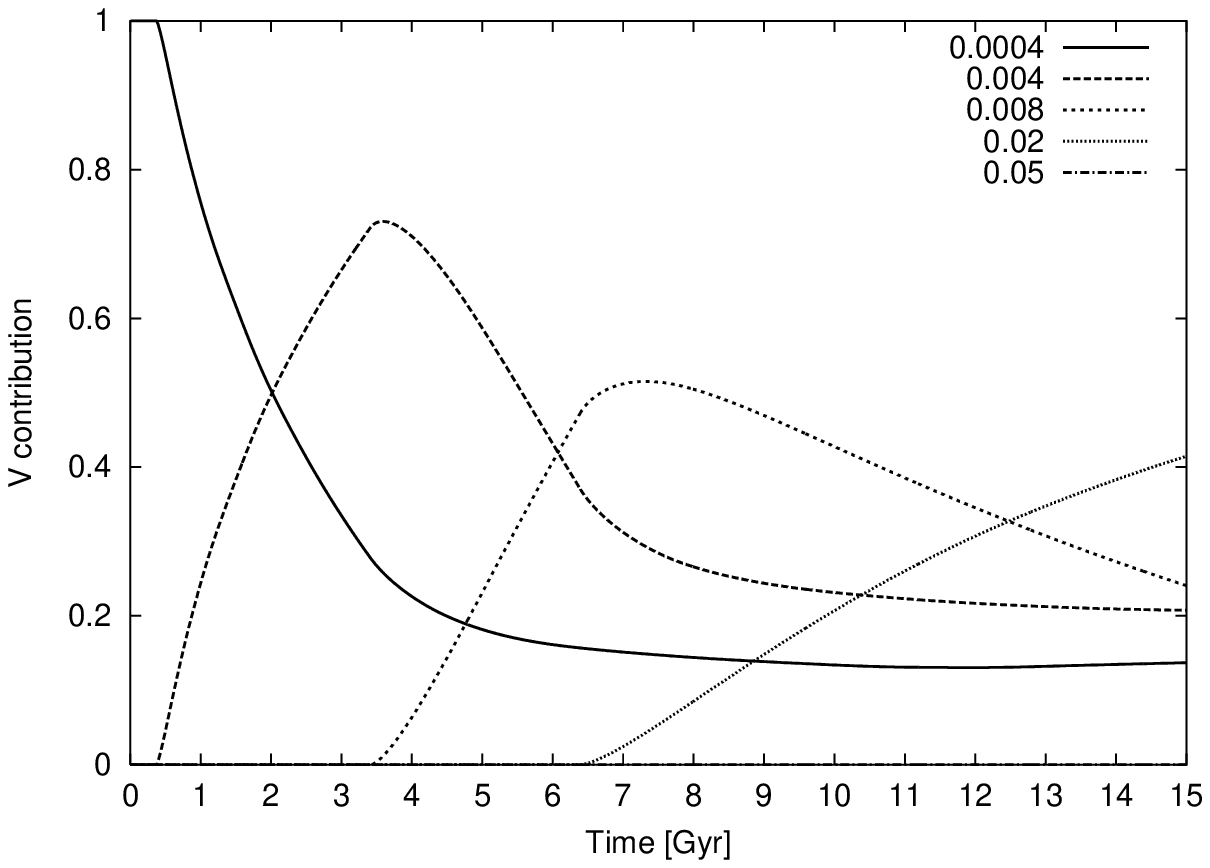}} 
\caption{Time evolution of the luminosity contribution to the V-band of the different 
metallicity subpopulations to the total luminosity for the E (top) and Sb (bottom) 
models} 
\label{ztime} 
\end{figure} 
 
\begin{figure} 
\resizebox{7.5cm}{!} {\includegraphics{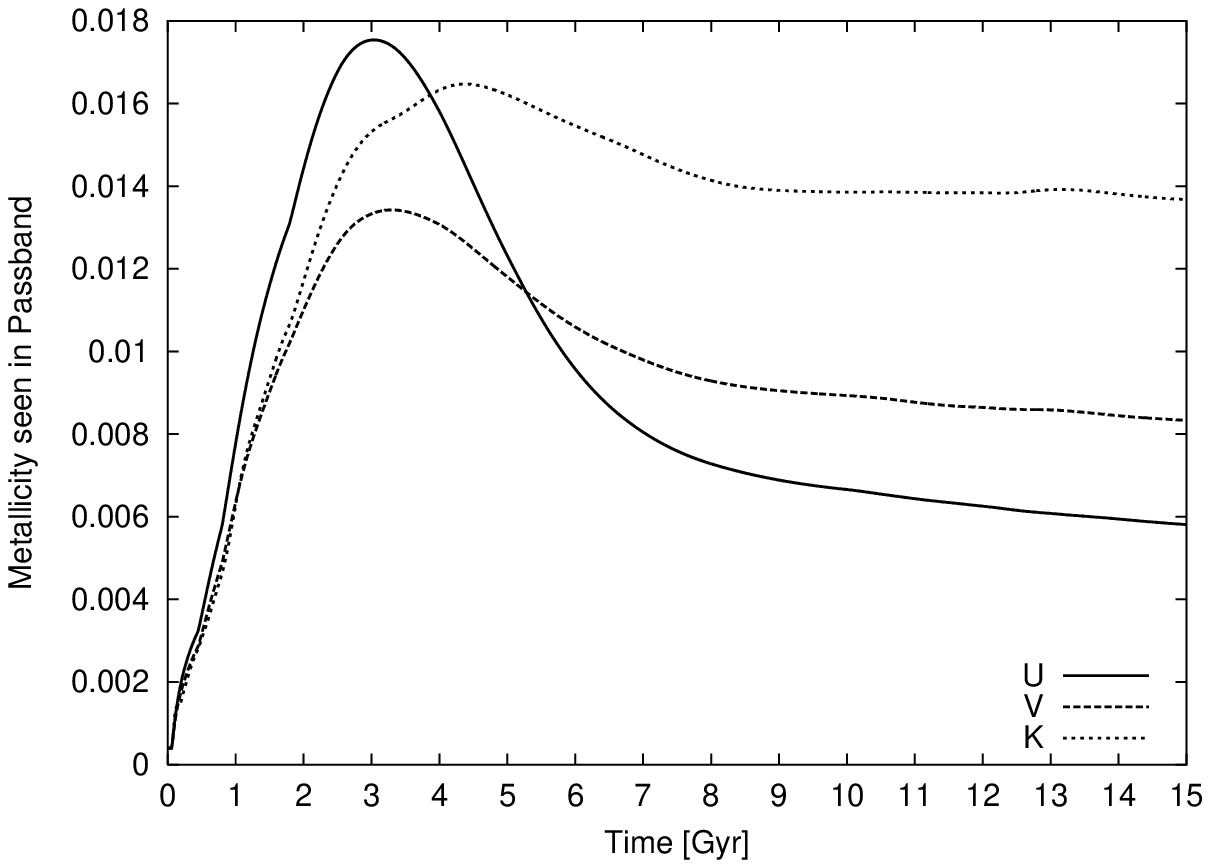}} 
\resizebox{7.5cm}{!} {\includegraphics{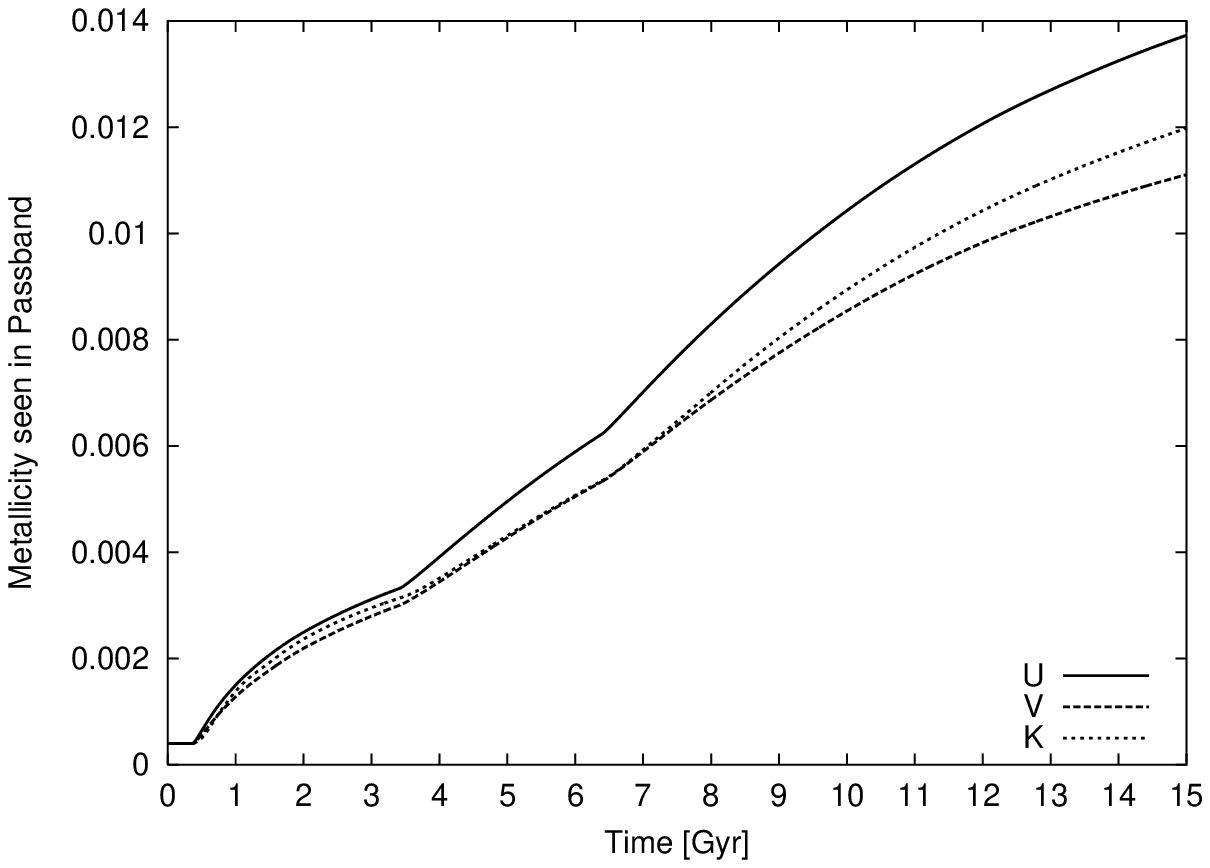}} 
\resizebox{7.5cm}{!} {\includegraphics{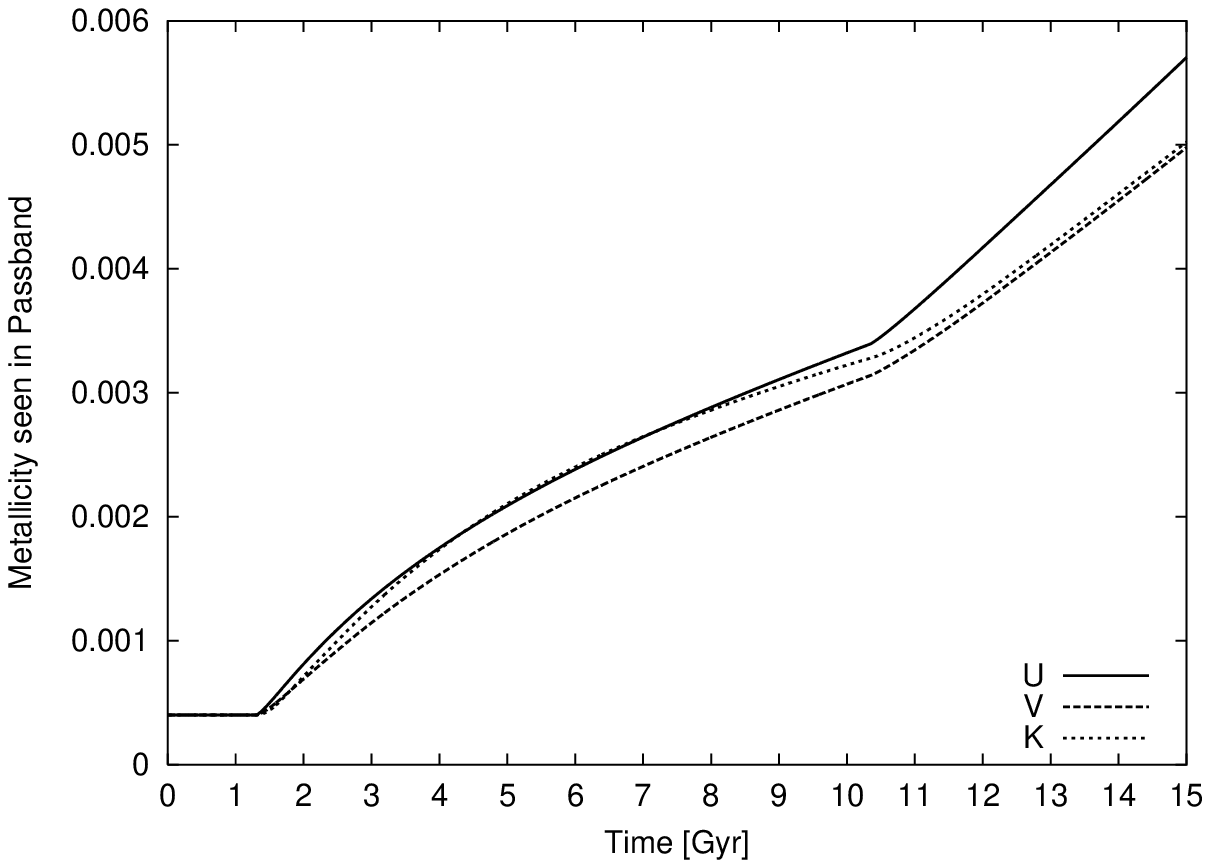}} 
\caption{Time evolution of the luminosity-weighted  stellar metallicities 
in U, V, and K (restframe) for different spectral types E, Sb, Sd (top to bottom).} 
\label{zlum} 
\end{figure} 
 
While Figs. \ref{hist12} and \ref{hist6} give the luminosity contributions of
different metallicity subpopulations   at ages of 12 and 6 Gyr, the time
evolution of the   
luminosity contributions of stellar subpopulations of different metallicities to
the V-band  is shown in Fig. \ref{ztime} for E and Sb galaxy models. 

It is seen that the broad stellar metallicity distribution of the E model is
already established at very young ages despite the present-day small stellar age
distribution.   Note in particular that despite its SFR declining rapidly on the
short timescale   of ${\rm t_{\ast} = 1}$ Gyr, our E model -- due to its
realistic stellar metallicity   distribution -- will differ significantly in its
spectrophotometric evolution from that   of any single metallicity single burst
model often used in the literature for the   interpretation of E galaxy
observations (see also Vazdekis \et\ 1996, 1997).

Over more than the last 50\% of its lifetime, the V-band light of
the elliptical model is  coming from stellar subpopulations of 4 different
metallicities (Z=0.0004 -- 0.02) at roughly comparable rates. The V-light of the
Sb-model is seen to have been dominated by stars with half-solar metallicity
during the second  half of its lifetime. Only very recently, i.e. at ages $\ge
10$ Gyr, solar metallicity stars gained importance while at early stages $< 6$
Gyr only stars with ${\rm Z\le 1/4~Z_\odot}$ where present.

Fig.~\ref{ztime} very clearly shows that at earlier   evolutionary times, as observed in
galaxies at high redshift, stars of lower and   lower metallicities become
dominant in the spectra of all spiral galaxies, though not for the E-model. And
this is, of course, not only   true in the V-band shown here, but over all
wavelengths.  

Fig.~\ref{zlum} shows the time evolution of the luminosity-weighted mean stellar 
metallicities, as defined in M\"oller \et\ (1998), in different passbands for E, Sb,
and Sd models. 
These luminosity weighted mean stellar metallicities in certain passbands are what
metallicity dependent absorption features in the respective wavelength range of the
integrated galaxy spectra are expected to measure.
As already indicated in Fig.~\ref{hist12} for an age of 12 Gyr  there
is a significant difference in the luminosity weighted mean stellar  metallicity in
various bands in the E-model and this is seen in Fig.~\ref{zlum} to have persisted for
all ages $\ge 7$ Gyr. The K-band shows the highest stellar metallicity of $<Z_K> \sim
0.014$, more than 50\% higher than that seen in V:  $<Z_V> \sim 0.008$. In the
rest-frame K the time evolution of the seen stellar metallicity is strongest. While it
hat rapidly reached a maximum of $<Z_K>_{max}\sim Z_\odot$ at an age around 3 Gyr it
tremendously decreases to a present $<Z_K>\sim 1/3~Z_\odot$ since then. The light
emitted in rest-frame U at a time around 3 Gyr, corresponding to a redshift ${\rm z\le1.5}$
is shifted to observer-frame I, so we expect to see much higher metallicities in high-z
ellipticals than in local ones.

Due to the longer timescales of SF the metallicity differences between different
passbands get much smaller towards later galaxy spectral types, as seen in
Fig.~\ref{zlum}, and their time evolution gets much slower and monotonic. Note the
scale differences among E, Sb, and Sd models.

%______________________________________________________________ 
 
\section{Evolutionary and Cosmological Corrections} 
 
All evolutionary and cosmological corrections published so far were calculated
with  models using solar metallicity input physics only (e.g. Bruzual \& Charlot
1993,   Poggianti 1997, Fioc et al. 1997).   Also still widely in use are the
k-corrections from Coleman \etal (1980) extracted from  observed template spectra,
although no evolutionary corrections are available in this case.    The wavelength
range of Colman \et\  observation (1400 -- 10000 \AA) limits the redshift range for
their k-corrections to z$\lesssim 1$ in U and to z$\lesssim 2$ in the R-band.

To cope with observations of very   high redshift galaxies up to ${\rm z \sim 5}$,
models are needed that include the far UV as well as evolutionary corrections.  

%______________________________________________________________ 
\subsection{Comparison with solar metallicity models}

In the following we will focus on the differences as a function of redshift between
our   chemically consistent models (cc models)  and models using solar metallicity
input physics only (${\rm Z_{\odot}}$ models).   To isolate the effects of the
chemically consistent treatment from those due to differences in the   codes of
various authors   and/or the particular sources of input physics chosen (Padova vs.
Geneva stellar   evolutionary tracks, observed stellar spectra vs. model atmosphere
libraries, etc.)   we ran ${\rm Z_{\odot}}$ models with our code and the same solar
metallicity input   physics as in our cc models. The comparison between cc and ${\rm
Z_{\odot}}$ models is   done for cosmological parameters ${\rm
(H_0,~\Omega_0,)~=~(65,~0.1)}$. Note that the attenuation is not included in this
comparison because it dominates the the e+k-corrections at high redshifts and would
mash the metallicity effect.  

In order to obtain agreement, after a Hubble time, with observed colours of the  
respective galaxy types (and, hence, with our cc model colours) the SFHs have to   be
slightly different in the models that do not contain the -- on average -- bluer   and
more luminous contributions of low metallicity stars. The ${\rm Z_{\odot}}$-E model
hence is to be described by an exponentially decreasing SFR with an e-folding   time
of ${\rm t_{\ast} \sim 2}$ Gyr as compared to ${\rm t_{\ast} \sim 1}$ Gyr for   the cc
E model. The ${\rm Z_{\odot}}$-Sb model needs ${\rm t_{\ast} \sim 9}$ Gyr   as
compared to ${\rm t_{\ast} \sim 7}$ Gyr for the cc Sb model.   And ${\rm t_{\ast} > 15
}$ Gyr for the  ${\rm Z_{\odot}}$ and the cc Sd model. 

\begin{figure} 
%\vspace{0cm} 
\centerline{\resizebox{7.0cm}{!}{\includegraphics{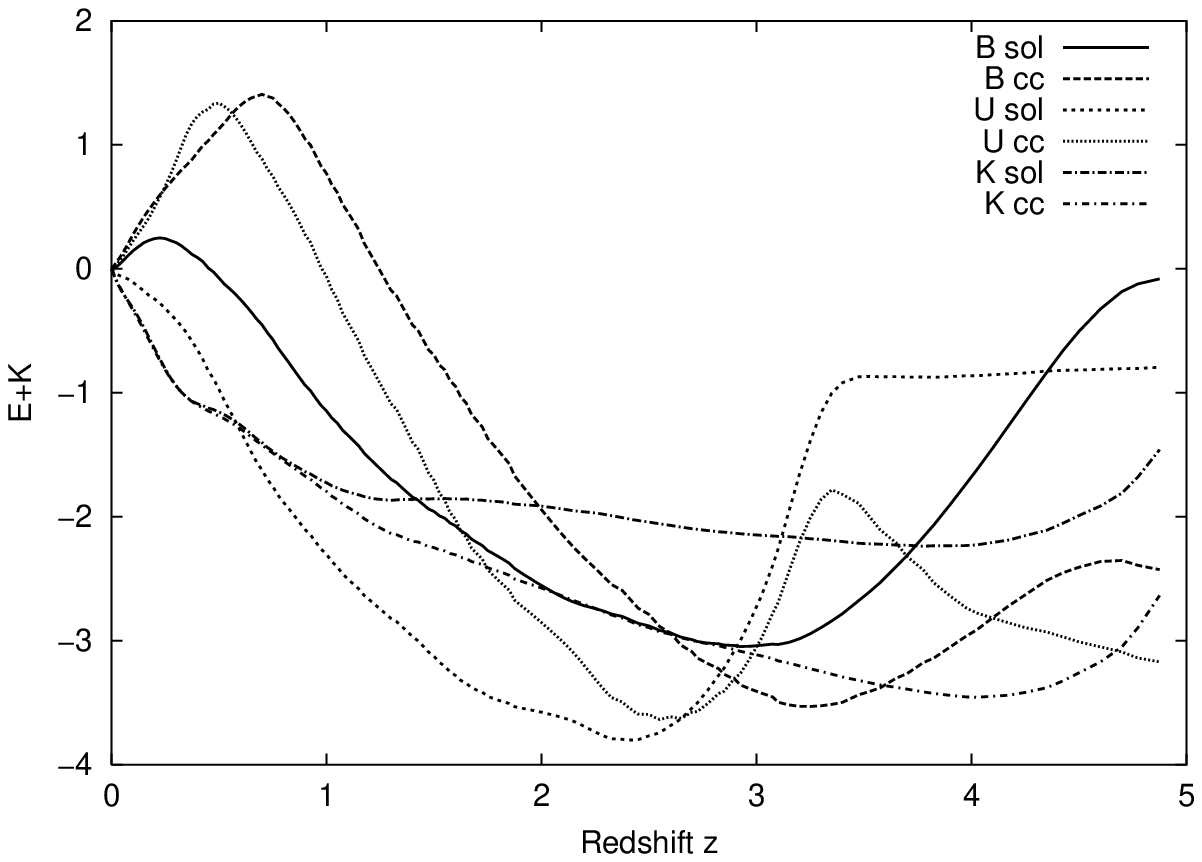}}} 
%\vspace{-0.5cm} 
\centerline{\resizebox{7.0cm}{!}{\includegraphics{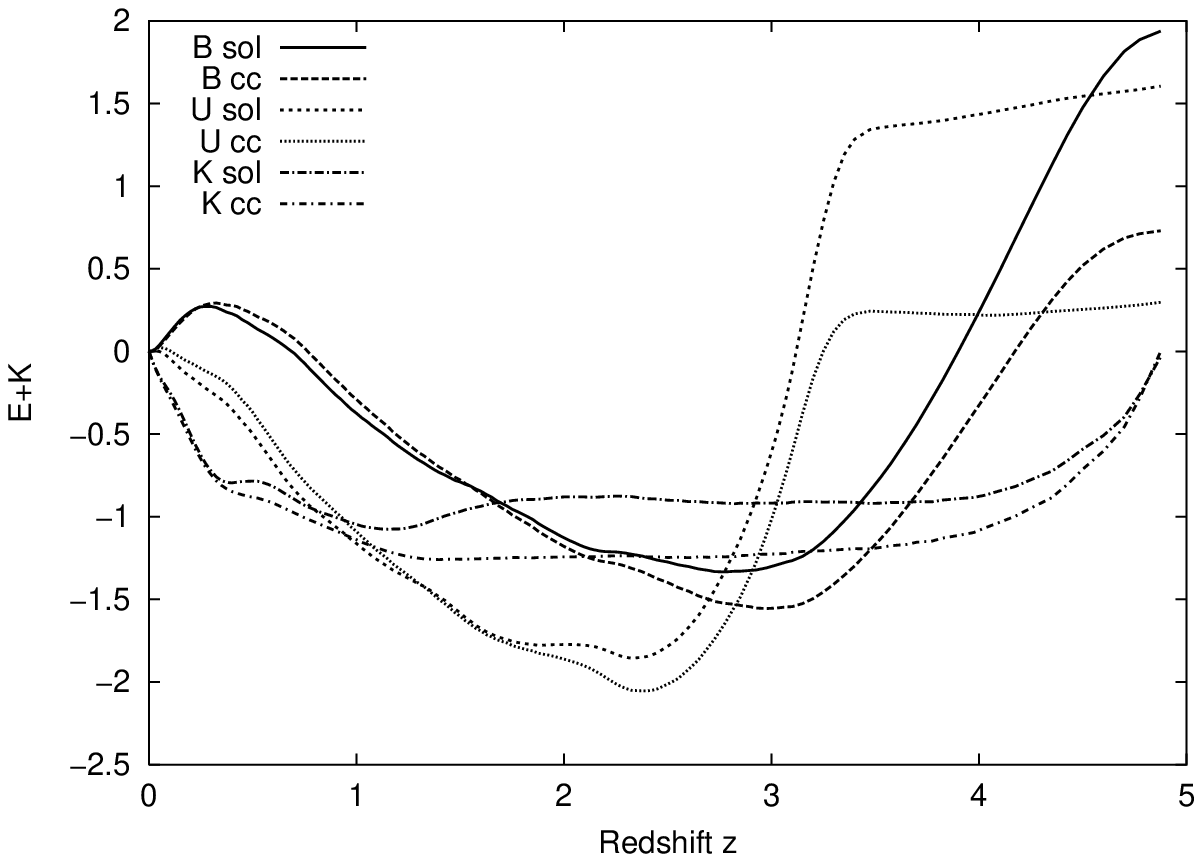}}} 
%\vspace{-0.5cm} 
\centerline{\resizebox{7.0cm}{!}{\includegraphics{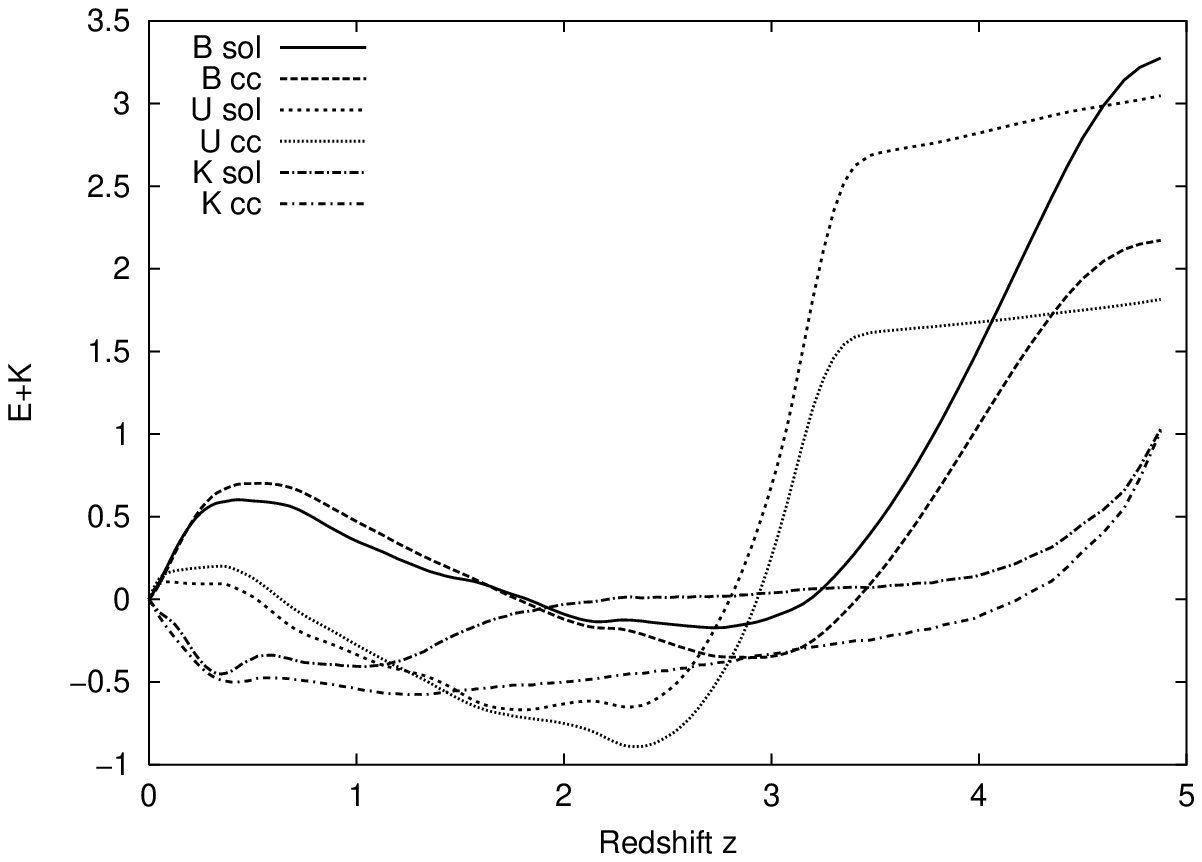}}} 
%\vspace{0cm} 
\caption{Comparison of the ${\rm (e+k)}$ corrections in cc and ${\rm Z_{\odot}}$
models for E (top), Sb (middle) and Sd (bottom) galaxies in U, B and K band  
${\rm (H_0,~\Omega_0,)~=~(65,~0.1)}$.} 
\label{ccek} 
\end{figure} 
 
\begin{figure} 
\resizebox{7.5cm}{!} {\includegraphics{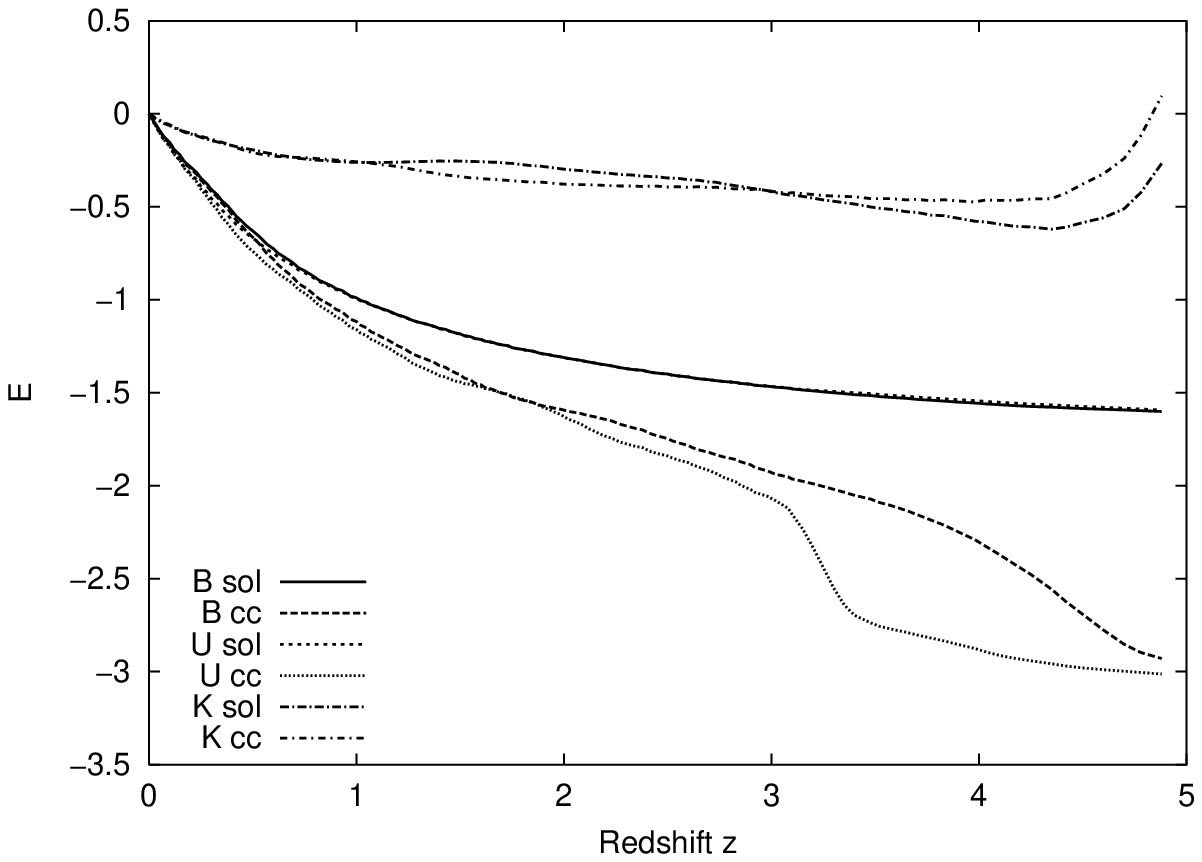}} 
\resizebox{7.5cm}{!} {\includegraphics{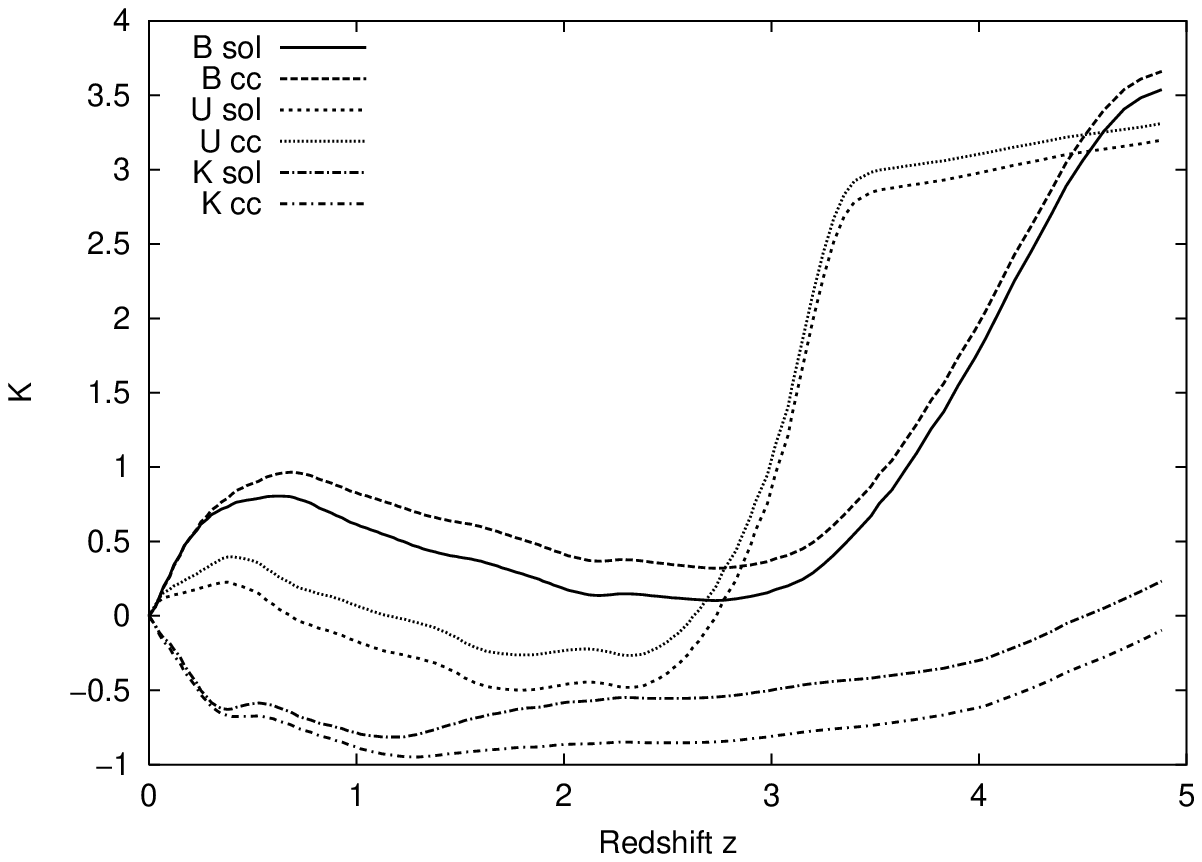}} 
\caption{Comparison of cc and ${\rm Z_{\odot}}$ models for Sb in the U-, B-, and K-band. Decomposition into evolutionary (top) and cosmological (bottom) corrections.} 
\label{ccecck} 
\end{figure}

Figs. \ref{ccek} and \ref{ccecck} show a comparison of e- and k-corrections in various
bands   as a function of redshift for   the cc and ${\rm Z_{\odot}}$ models.   The
most obvious difference between cc and ${\rm Z_{\odot}}$ models is seen   for the E
model at short wavelengths. While for redshifts ${\rm z \lta 1.2}$,   the cc ${\rm
(e+k)}$ corrections in U and B are positive, making the apparent   magnitudes of
ellipticals fainter, those of the ${\rm Z_{\odot}}$ model get increasingly negative  
from ${\rm z = 0}$ through ${\rm z \sim 2.5}$ in U and to ${\rm z \sim 3}$ in B,
respectively (cf. Fig. \ref{ccek}),   making ellipticals apparently brighter 

The maximum difference for the E models is as much as $\sim 3$ mag in U and
$\sim 2$ mag in the B-band at ${\rm z \sim 1}$. 
Hence, {\bf cc models with a SFR decreasing exponentially with an e-folding
time of 1 Gyr   predict a smaller number of ellipticals to be expected at  
${\rm\bf z \lta 2.5}$ in magnitude limited samples.} At redshifts $\gta 2.5$
the  situation is reversed and the cc models appear more luminous. But our
simple classic E model may not applicable at this redshift range.
At longer wavelengths, e.g. K, differences are very small by redshift $\sim 1$
and increase to $\sim 1.5$ mag by ${\rm z=4}$.  

For ${\rm z \lta 2}$ the differences in U and B between our Sb and Sd cc and
the respective ${\rm Z_{\odot}}$ models are small. At the maximum the cc model
has a $\sim 0.1$ mag higher correction. At redshifts higher than 2 the
differences of the models increases and the ${\rm Z_{\odot}}$ models have
higher corrections than the cc models. The difference grow up to 1 -- 1.5 mag
in  the U and B band at high redshift. For the K-band the  ${\rm Z_{\odot}}$
models  have always an higher e+k-correction than the cc models with a maximum
difference of 0.4 -- 0.5 mag at ${\rm z\sim 2}$.

Hence, {\bf on the basis of our cc models we expect a slightly smaller number of
late-type spirals  by redshifts ${\rm z \lta 2}$ to show up in magnitude
limited surveys (U and B) as compared to ${\rm Z_{\odot}}$ model predictions.
At ${\rm z\gta 2}$ we expect more late-type spirals as compared to the ${\rm
Z_{\odot}}$ models } 

Fig. \ref{ccecck} presents the decomposition of the ${\rm (e+k)}$-corrections
shown in Fig. \ref{ccek} into   the evolutionary   and cosmological corrections ${\rm
e_{\lambda}(z)}$ and ${\rm k_{\lambda}(z)}$ for the   cc and ${\rm Z_{\odot}}$ Sb
model in the U-, B-,and K-bands.  

Let us first look at the e-correction. In the ${\rm Z_{\odot}}$ model B-band
e-corrections   become increasingly negative from ${\rm z = 0}$ to ${\rm z \sim
2}$ and remain constant around  ${\rm e_B \sim -1.5}$ mag all through ${\rm z > 4}$. In
the cc model ${\rm e_B}$   decreases faster to ${\rm e_B}\sim -3$ at ${\rm z \sim 5}$.
Without the strong compensating effect of the k-correction the   strongly negative
${\rm e_B}$-correction would make the cc Sb model appear much brighter at   ${\rm
z \gta 1.5}$ than the ${\rm Z_{\odot}}$ model. At ${\rm z \gta 1.5}$ the galaxies
in our   cosmology have ages $\lta 2.5$ Gyr and stars of low metallicities Z$\sim
0.0004 - 0.004$  dominate the light in B. The integrated spectrum still shows a
high flux in B so shortly after the major SF epoch. The behaviour of the ${\rm
e_U}$ -correction is very similar to ${\rm e_B}$.
In K the evolutionary corrections are ${\rm e_K} \lta -0.5$ through ${\rm z\sim 4}$ for
both, the cc and ${\rm Z_{\odot}}$ Sb-models. 

\begin{figure}  %\vspace{0cm} 
\resizebox{7.5cm}{!} {\includegraphics{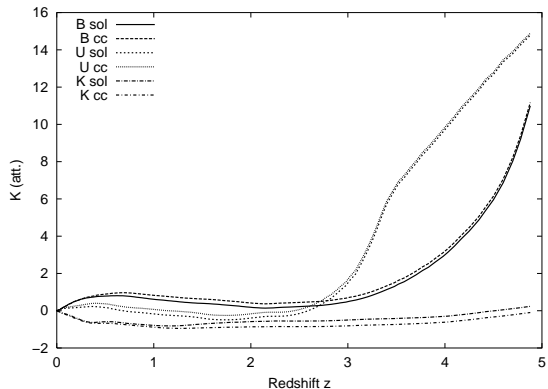}}  
\caption{K-correction --including the attenuation-- for the Sb-model. }  
\label{katt}  
\end{figure}  

For the k-corrections, the situation is quasi reversed. Till ${\rm z \sim 0.5}$ the
${\rm k_B}$-correction for the cc Sb model increases to ${\rm k_B}\sim 1$ and then it
slightly falls to  ${\rm k_B}\sim 0.4$ at ${\rm z \sim 3}$. Beyond ${\rm z=3}$, ${\rm
k_B}$ increases rapidly to 3.7 at ${\rm z\sim 5}$. For the ${\rm Z_{\odot}}$ model
${\rm k_B}$ follows a trend similar to that of the cc model but with an offset of ${\rm
\sim -0.3}$. The ${\rm k_U}$-corrections show a redshift evolution similar to that of
${\rm k_B}$. However, the ${\rm k_U}$ is smaller than ${\rm k_B}$ at lower redshifts
($\la 0.4$) and negative in the redshift range  ${\rm z\sim 1.3 - 2.6}$. From ${\rm
z\sim 2.6 - 3.5}$ the ${\rm k_U}$-correction increases from 0 to $\sim 3$ as the
Lyman-break gets redshifted into the U-band. Beyond ${\rm z=3.5}$ the Lyman-break is
shifted out of the U passband and hence the curve flattens. The difference between the
cc and the ${\rm Z_{\odot}}$ model is again an offset of ${\rm \sim -0.3}$. The K-band
${\rm k}$-correction for the cc model is always negative. It drops down to ${\rm
k_K}\sim -0.9$ at redshift ${\rm z\sim 1.3}$ and rises to ${\rm k_K}\sim 0$ at ${\rm
z=5}$. Until ${\rm z=1}$ the difference between cc and ${\rm Z_{\odot}}$ models is
small ($\lta 0.1$). At redshifts $\ga 1.5$ the cc model has an offset in
${\rm k_K}$ of $\sim 0.3 - 0.4$.

The k-corrections and the e+k-corrections shown before do not include the
effect of attenuation to highlight the differences between the cc and the
${\rm Z_{\odot}}$ models. Fig \ref{katt} shows the k-corrections for the Sb
model including the attenuation (cf. Fig.~\ref{ccecck}). It is seen that the
attenuation is the dominant effect beyond ${\rm z=3}$ in U and ${\rm z=4}$ in
B. 

%______________________________________________________________ 
\subsection{Comparison with earlier work}  

\begin{figure} 
\centerline{\resizebox{8.5cm}{!}{\includegraphics{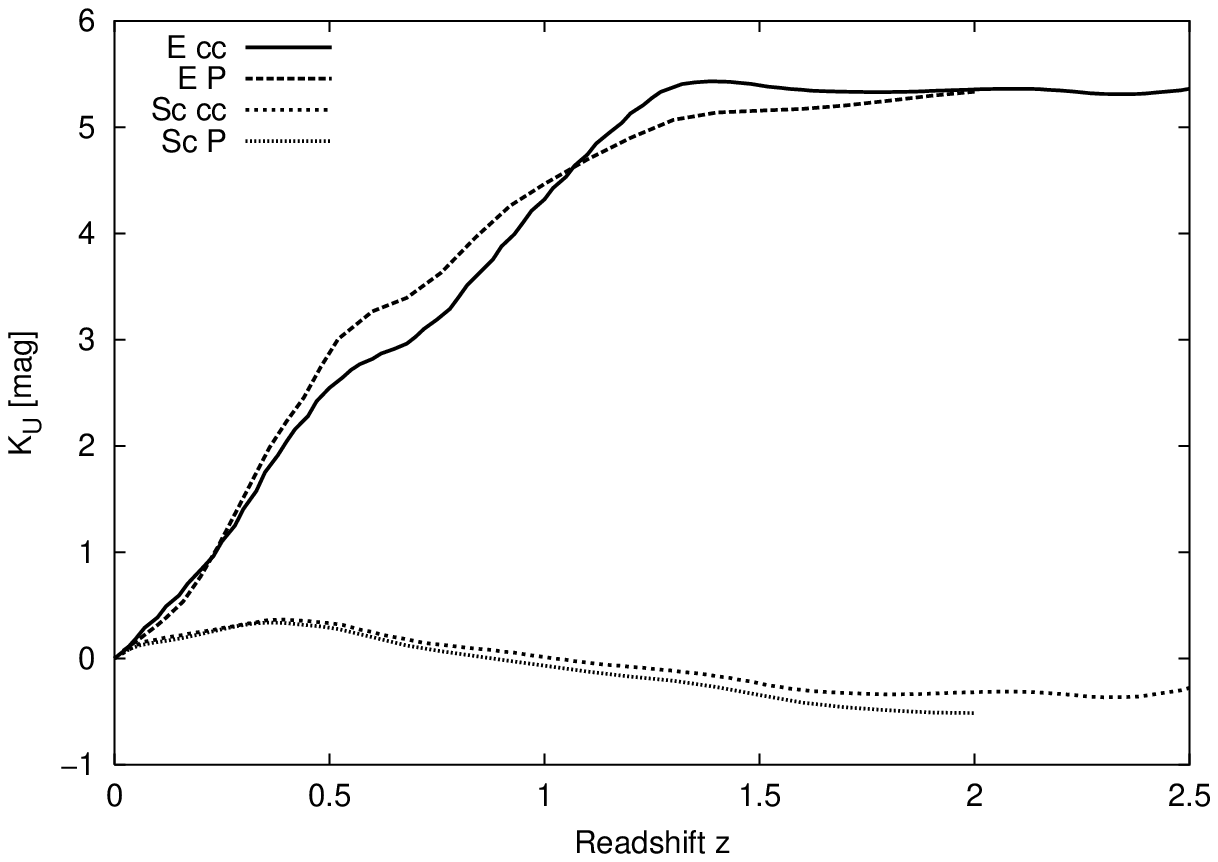}}} 
\vspace{0cm} 
\centerline{\resizebox{8.5cm}{!}{\includegraphics{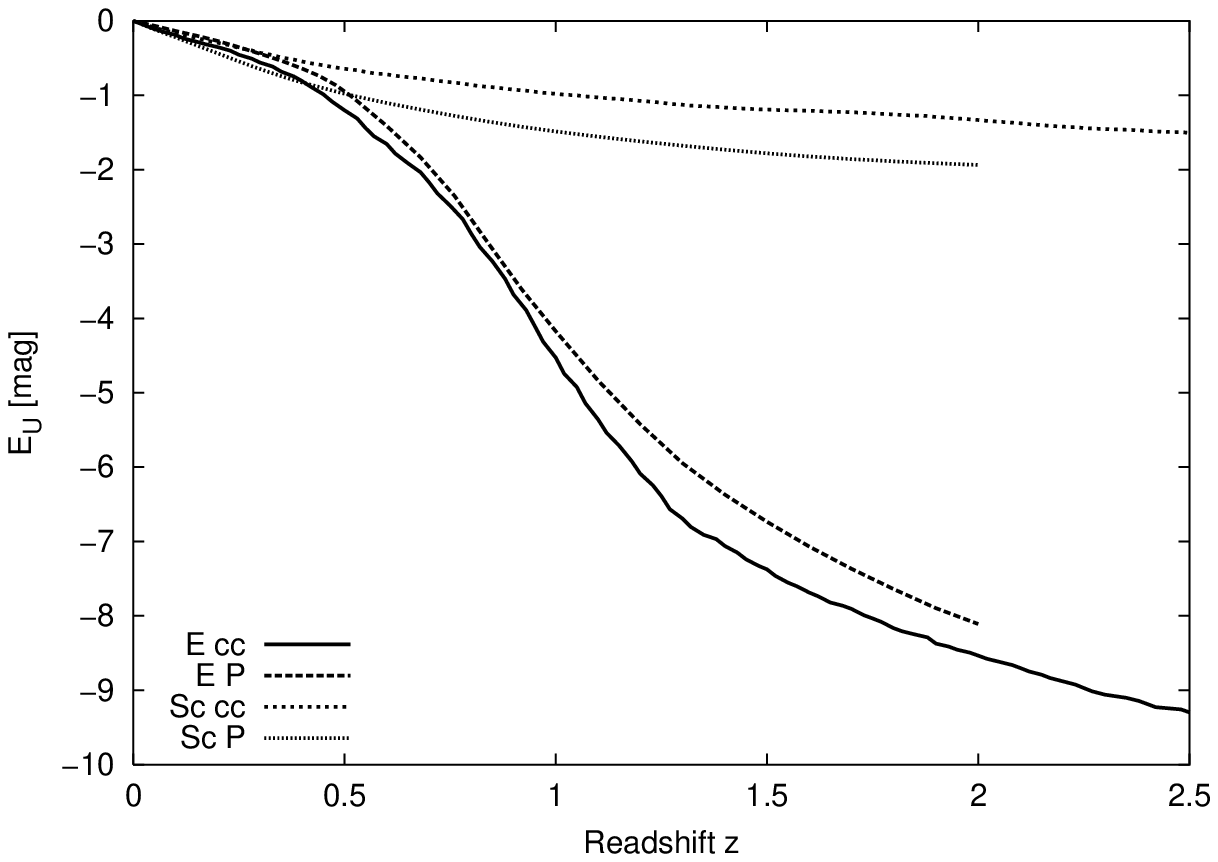}}} 
\vspace{0cm} 
\caption{Comparison between our cc models and the solar models of Poggianti
(1997) (P). The top panel shows the k-corrections and  the bottom panel the
e-corrections for E and Sc models in U. Note that we compare none-attenuated 
models in Poggianti's cosmology ($H_0,\Omega_0$)=(50,0.5).} 
\label{ekccvsp}
\end{figure}

\begin{figure} 
\centerline{\resizebox{8.5cm}{!}{\includegraphics{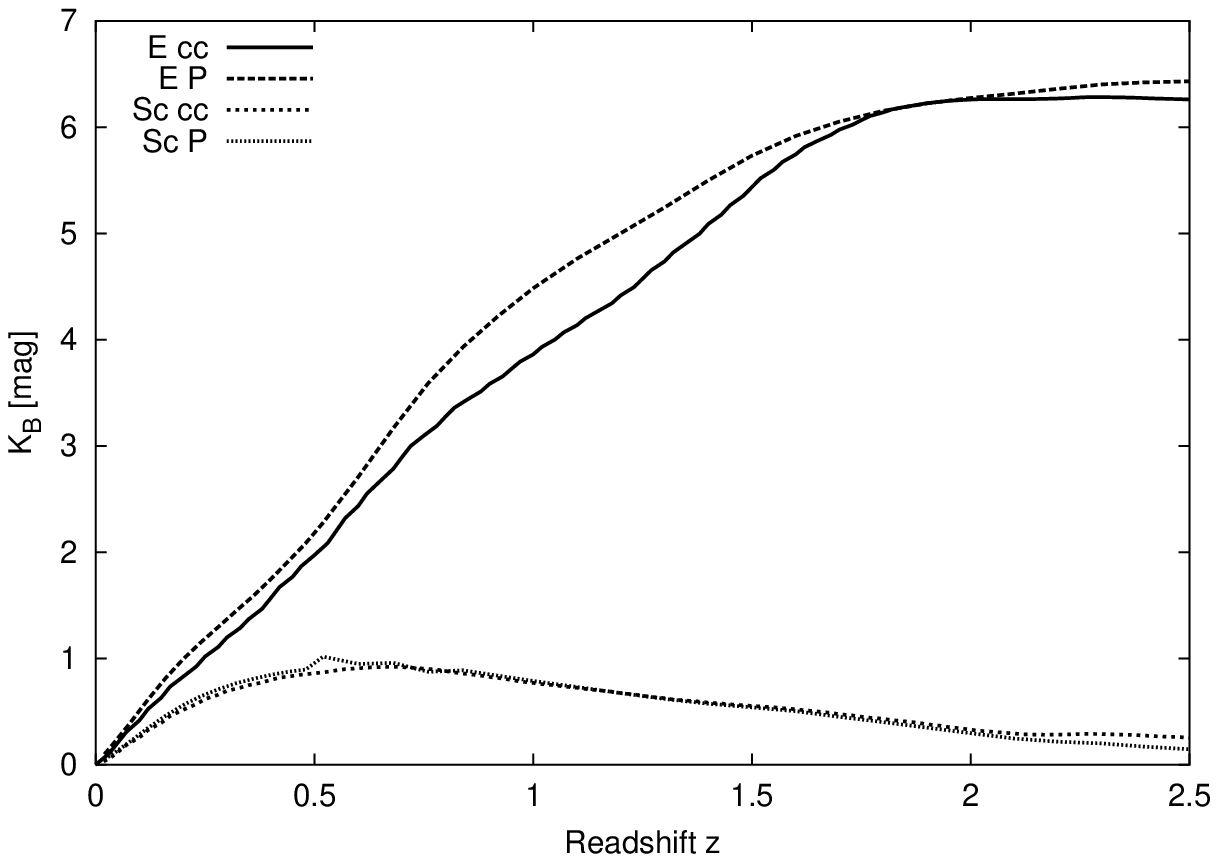}}} 
\vspace{0cm} 
\centerline{\resizebox{8.5cm}{!}{\includegraphics{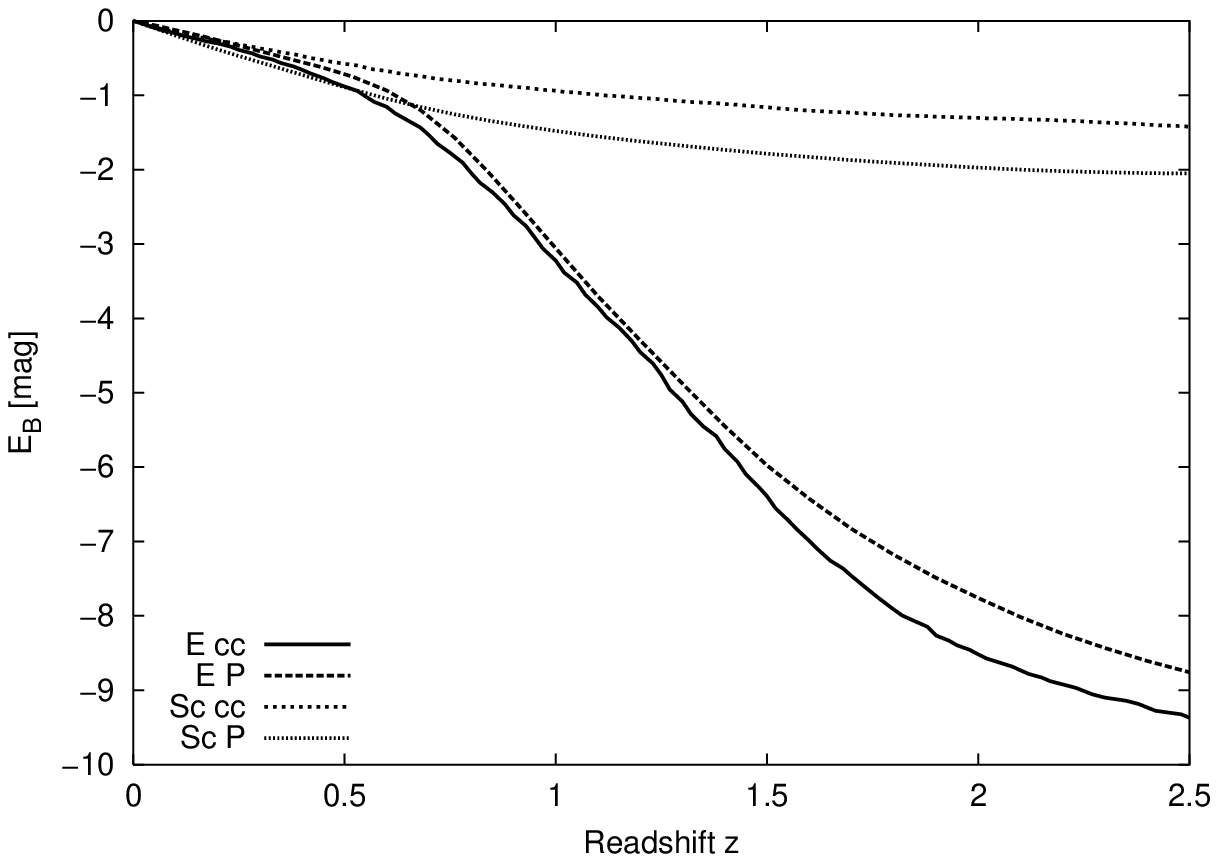}}} 
\vspace{0cm} 
\caption{Same like Fig. \ref{ekccvsp}, but now in the B-band} 
\label{ekccvsp1}
\end{figure}

We compare both our model e- and k-corrections to those of Poggianti (1997).
She calculated the e and k correction for solar metallicity models. Note that
her models do not include the attenuation by   intergalactic HI and -- due to a
shorter wavelength coverage -- do not allow for cosmological corrections in U
for ${\rm z > 2}$ and in B for ${\rm z > 2.5}$. So we compare with our non
attenuated cc models at this point. We also calculate the e-correction for the
cosmology given by Poggianti (1997) for this comparison (${\rm
H_0=50,~\Omega_0=0.5}$). 
Figs. \ref{ekccvsp} and \ref{ekccvsp1} show the comparison between our E and Sc cc
models with the respective models form Poggianti (P) in the U- and B-band. Note the
good agreement of both models. 

\begin{table}\centering 
        \caption{Evolutionary and cosmological corrections for ($H_0,\Omega_0$)=(65,0.1). 
                 The complete version of the table is given on our homepage.} 
        \begin{sf} 
        \begin{tabular}{|r|*{2}{r}|*{2}{r}|*{2}{r}|*{2}{r}|}\hline 
        \multicolumn{9}{|c|}{\rule[-2mm]{0mm}{6mm} Sb model} \\ \hline 
         $z_r$& e$_U$& k$_U$&  e$_B$&  k$_B$& ... &  & e$_K$& k$_K$ \\ \hline 
          0.0 & 0.00 & 0.00  &  0.00 & 0.00 & ... &  & 0.00 &  0.00 \\ 
          0.1 & -0.17 & 0.18 & -0.16 & 0.26 & ... &  & -0.05 & -0.22 \\ 
          0.2 & -0.33 & 0.26 & -0.29 & 0.52 & ... &  & 0.62 & -0.42 \\ 
          0.3 & -0.48 & 0.34 & -0.41 & 0.71 & ... &  & 0.52 & -0.61 \\ 
          .   &   .   &      &       &      & ... &  &      &    \\ 
          :   &   :   &      &       &      & ... &  &      &    \\ 
        \hline 
       \end{tabular} 
       \end{sf} 
\end{table}

%------------------------------------------------------- 
 
\section{Redshift evolution of apparent magnitudes} 
 
\begin{figure} 
\centerline{\resizebox{8.5cm}{!}{\includegraphics{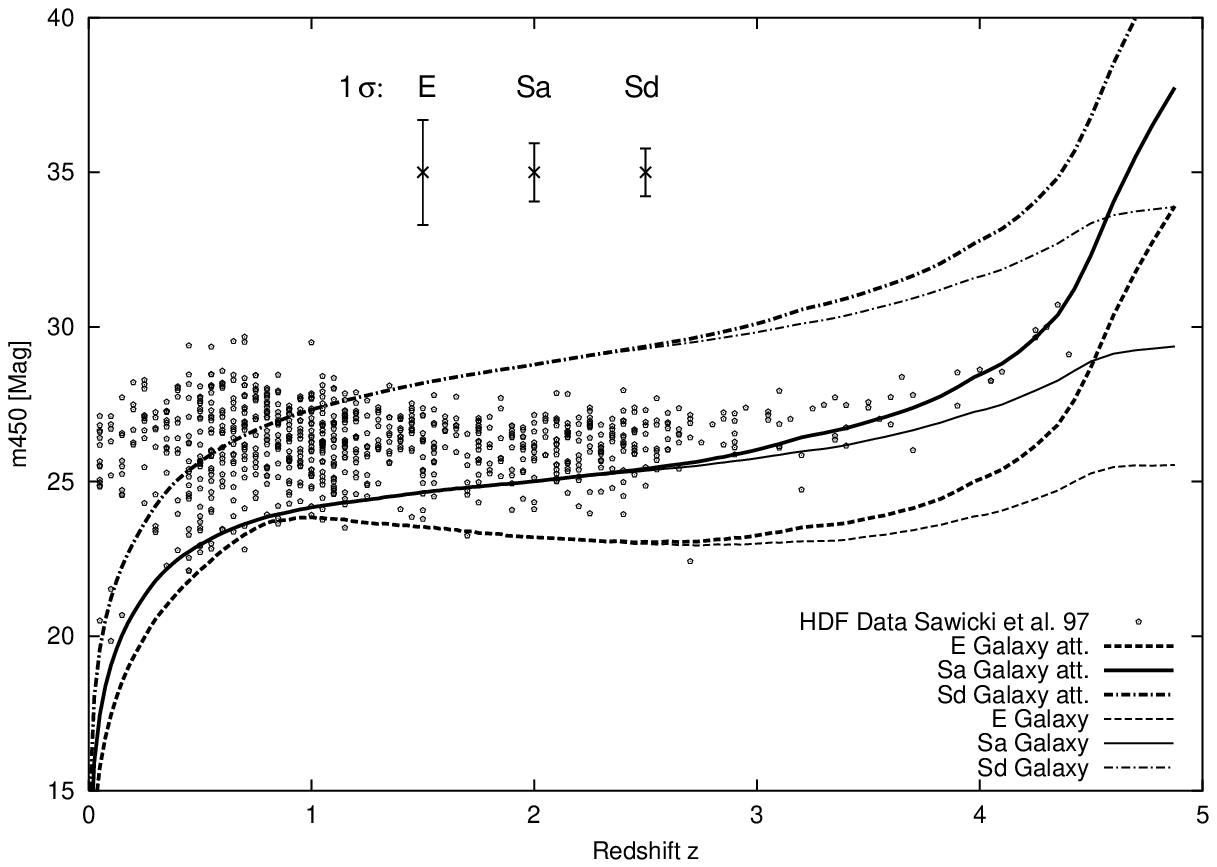}}} 
\vspace{0cm} 
\centerline{\resizebox{8.5cm}{!}{\includegraphics{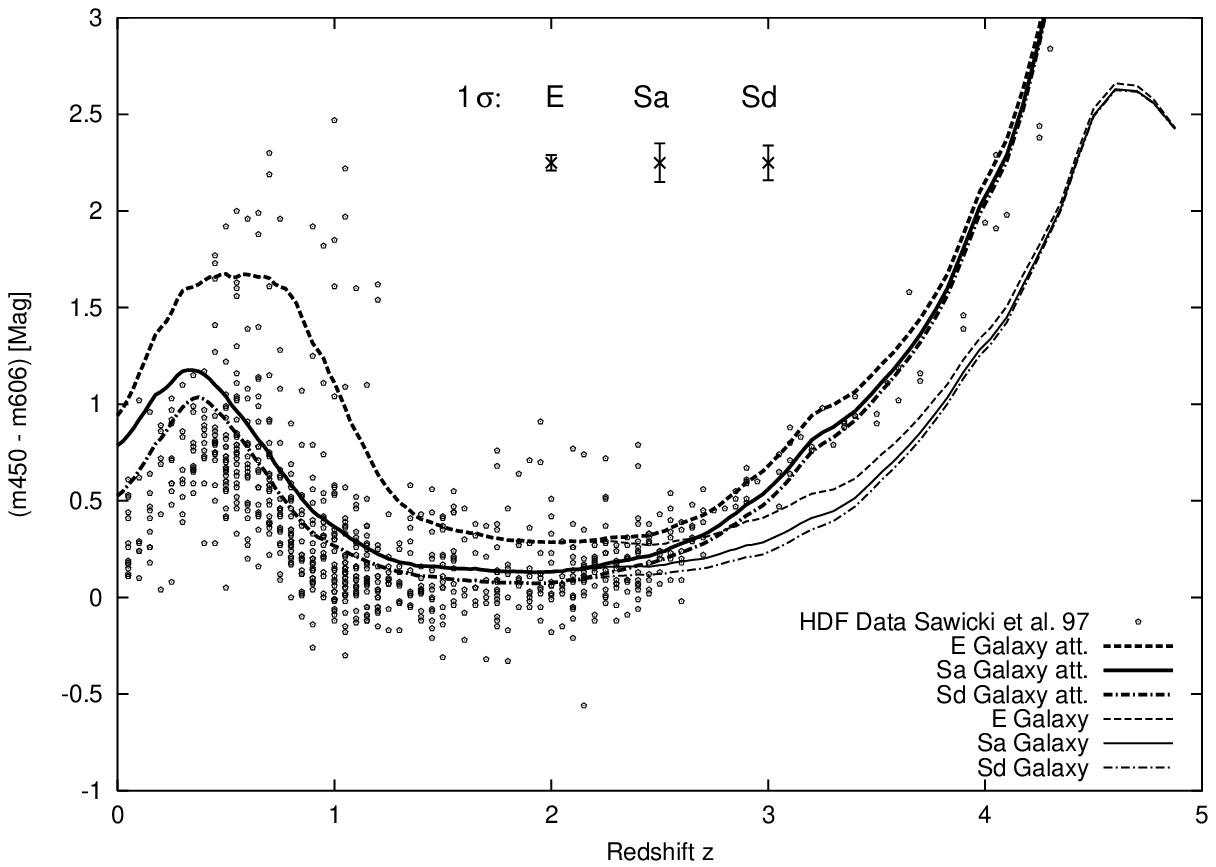}}} 
\vspace{0cm} 
\caption{Redshift evolution of B-band luminosity (m450) and from the B--V colour (m450-m606) compared 
to  the HDF galaxies photometric redshift catalogue of Sawicki et al. (1997). Vertical bars show
the $1\sigma$ luminosity and colour ranges of local galaxies} 
\label{AppMag}
\end{figure} 
 
The redshift evolution of apparent magnitudes in U,B,V,${\rm R_C}$,${\rm
I_C}$,J,H,K and HST filters is given on our homepage.  

In Fig. \ref{AppMag}, we show the redshift evolution of apparent magnitudes for
the E, Sa, and Sd model in in HST m450-band ($\sim$Johnson B) and (${\rm m450 -
m606}$) ${\rm \sim(B - V)}$ colours and compare to the photometric redshift catalogue for the
HDF galaxies (Sawicki et al. 1997). For each model the evolution is plotted with
(thick lines) and without (thin lines) attenuation.   

Before we present a very first and preliminary comparison of cc models with LBG data
we recall   that ${\rm 1^{st}}$ our models refer to integrated properties of galaxies
and ${\rm 2^{nd}}$ our   model galaxy types -- being described by SFHs or SF
timescales -- refer to a spectral and not to a   morphological classification of
galaxies. The question how far back in time the locally   observed 1-to-1
correspondence between spectral and morphological galaxy types remains valid is
another open question.

The E model becomes apparently fainter at redshifts ${\rm 0 < z \lta 1}$ (${\rm
m450\sim 24}$mag), it then starts getting brighter again as it approaches its active
star formation phase. At ${\rm z \sim 3}$ the attenuation comes into play and makes
the model get fainter rapidly. By ${\rm z = 5}$ the difference between the attenuated
and the non-attenuated models increases to $\sim 10$ mag. 

The Sa and Sd models become fainter at redshifts ${\rm 0 < z \lta 1}$, with ${\rm
m450\sim 24}$mag for Sa and ${\rm m450\sim 27}$mag for Sd at ${\rm z\sim 1}$.
Thenceforward both spiral models get moderately fainter until the attenuation comes
into play at ${\rm z \sim 3}$.

In comparison to the HDF data from Sawicki et al. (1997), the Sa model
follows the bright limit familiar well. The Sd model gets fainter than the
observational limit of ${\rm m450 \la 29}$mag beyond ${\rm z = 2}$ while the
classical initial collapse E model is too bright at ${\rm z \gta 2}$ to be
comparable with the data. The lack of luminous galaxies at redshifts lower
than 0.5 is the selection effect explicitly intended by the choice of the
HDF.   The lack of galaxies as luminous as our classical E model, however is
real and not due to any selection or bias. But on the basis of the
luminosity we can not determine the role of attenuation. To do this
observations ${\rm z \gta 4.5}$ are needed.

More information is provided by the colours of galaxies.  For the (${\rm m450
- m606}$) colour evolution it is seen that for   redshifts ${\rm z \ge 2 }$
the E model gets as blue as the spirals. This is partly due to   the strong
light contributions from low metallicity stars and partly due to   the youth
of its stellar population (age $\sim 1.6$ Gyr at ${\rm z \sim 2}$). In the
redshift range ${\rm 0.25\la z \la 0.75}$ the E model is very red ${\rm (m450
- m606)\sim 1.7}$. The colour difference between the Sa and Sd models is
small, $\la 0.3$. The models follow the data well, but there are a lot of
very blue objects which cannot be explained by our models of undisturbed
galaxy evolution.    

At redshifts ${\rm z\ga 3}$ the models with attenuation fit the data mutch
better than models without attenuation.  While at ${\rm z\la 1}$ a large number
of faint and blue galaxies are seen in the HDF,bluer and fainter than  our
late-type spiral models, the bulk of galaxies at ${\rm 1\la z\la 3}$ are well
compatible wit our normal spiral progenitor models. 

A detailed modelling and interpretation of   these {\bf L}yman {\bf B}reak {\bf
G}alaxies ({\bf LBG}s) is beyond the scope of the present paper.   It will be the
subject of a forthcoming paper including a much larger number of LBGs   both with
spectroscopic and photometric redshifts (see M\"oller \et\ 2001b for first and
preliminary results). This investigation will use the full multi-colour information
available for HDF and other deep field galaxies.

%______________________________________________________________ 
\section{Conclusions}\label{conclude} 
 
Our evolutionary synthesis model is now chemically consistent   with respect to both
the  spectrophotometric and the chemical evolution.  This gives the possibility  for a
detailed study of the age and metallicity distributions in the composite stellar
populations of galaxies. In the framework of simplified closed-box models where the
different spectral types of galaxies E, Sa, Sb, Sc, and Sd are described by their
respective appropriate star formation histories we analyse   how the presently
observed average stellar abundances and abundance distributions have evolved in
time.  

While -- with somewhat different star formation histories -- models using input
physics for one single metallicity (e.g. solar)  only can also be brought into
agreement with locally observed galaxy colours and spectra, the evolution with redshift
is significantly different.  

Before analysing galaxy data out to high redshifts  we made sure that our
chemically consistent models correctly describe  the integrated properties of
galaxy types E, Sa--Sd   in their spectrophotometric appearance from the UV
through the NIR, their average ISM abundances,   the metallicity of their stellar
populations, their   gas content, present-day star formation rates   and the
chemical abundances of various elements.  

This large number of observables allows to tightly constrain the only free parameter  
of our closed-box 1-zone models: the time evolution of their star formation rate or
their star formation history. We show how the stellar metallicity distribution in
various galaxy types build up with time to yield after $\sim 12$ Gyr agreement with
stellar metallicity distributions observed in our and other local galaxies. We also
presented the time evolution of the luminosity-weighted mean stellar metallicities of
different galaxy types in various bands. It is these luminosity-weighted metallicities
that are expected to be measured by metal-sensitive absorption features observed in
the integrated light in the respective passbands.

For the spectral galaxy types E, Sa to Sd, we give the spectral energy
distributions  over a wide wavelength range (9.09 - 160000) nm  in their time
evolution from ages of 1 Gyr up to 15 Gyr.  

Any desired set of filter response functions can directly be used to calculate
from these spectra   the time and redshift evolution of galaxy luminosities,
colours, evolutionary and cosmological   corrections and apparent magnitudes (for
any desired cosmological model).  

In comparison with models using solar metallicity input physics only, we discuss
the effects   of the inclusion of subsolar metallicity stellar subpopulations.  
In chemically consistent models E galaxies appear significantly fainter to
redshifts   ${\rm z \sim 2.5}$ as compared to solar metallicity models. A smaller
number of ellipticals is thus expected to contribute to magnitude limited
samples.  

Chemically consistent spiral models, on the other hand, appear brighter at ${\rm z
\gta 2}$   than they would if only solar metallicity input physics were used.
Hence, we expect a larger   number of intermediate and late-type spirals from
${\rm z \gta 2}$ to show up in magnitude limited surveys.  

We present a large grid of evolutionary and cosmological corrections as well as
apparent magnitudes and colours in various filter systems (Johnson, HST, ...) from UV
to NIR including the effect of attenuation by intergalactic HI for galaxy  types E and
Sa to Sd using cosmological parameters   ${\rm (H_0,~\Omega_0)~=~(65,~0.1)}$ with a
redshift of  galaxy formation assumed to be ${\rm z_f = 5}$. 

Models and results our chemically consistent chemical evolution models in terms of a
large number of individual element abundances in the ISM of various spiral types were
presented in Lindner \et\ (1999) and used for the interpretation of Damped Ly$\alpha$
Absorbers. 

A very first and preliminary comparison with the redshift evolution of HDF galaxies
with photometric redshifts from Sawicki \et\ (1997) indicates that their luminosities,
colours and SFRs are well compatible with those  of our normal spiral models over the
redshift interval from ${\rm z\sim 0.5}$ all through ${\rm z >4}$.  

A detailed and extensive comparison of our chemically consistent model results with the
full set of observed colours and luminosities of high redshift galaxies (e.g. Lyman Break
Galaxies) will be presented  in a forthcoming paper.  

In the framework of chemically consistent models   it is also possible to include the
effects of dust  in a largely consistent way, tying the amount of dust to the evolving
gas content and metallicity.  Stellar track based chemically consistent models with
dust will be presented in a companion paper by M\"oller el al. ({\sl in prep.}, see
M\"oller \et\ 2001a, c for first results). 

\appendix

\section{Available data} 

The following data are available at our homepage {\sl http://www.uni-sw.gwdg.de/$\sim$galev/ccmodels/}:
e- and k-corrections, apparent magnitudes, spectra as function of redshift, and spectra as
function of evolution time given in restframe. These data are available for all Models (E-Sd).
Please see the README file for detailed information and file format. 

%______________________________________________________________ 
\begin{acknowledgements} 
  This work was partly supported by the Deutsche Forschungsgemeinschaft (DFG) 
  grant Fr 916/10-1. 
  We thank P. Madau for providing us with the attenuation functions 
  and T. Lejeune for sending us the corrected version of the  
  stellar atmosphere models. 
  
  We thank our referee, Dr. L. Carigi, for a insightful and constructive report.  
  
\end{acknowledgements}


\begin{thebibliography}{} 
 
\bibitem[1978]{aaronson} Aaronson, M., 1978, ApJ 221, L103 
 
\bibitem[Anders \& Fritze-v.~Alvensleben(2003)]{2003A&A...401.1063A} 
Anders, P.~\& Fritze-v.~Alvensleben, U.\ 2003, \aap, 401, 1063 

\bibitem[1986]{arimoto} Arimoto, N., Yoshii, Y., 1986, A\&A 164, 260 

\bibitem[Bertelli et al.(1994)]{1994A&AS..106..275B} Bertelli, G., Bressan, 
A., Chiosi, C., Fagotto, F., \& Nasi, E.\ 1994, \aaps, 106, 275  

\bibitem[1988] {bessel} Bessel, H.L., Brett, J.M., 1988, PASP 100, 1134 
 
\bibitem[1994] {bressan4} Bressan, A., Chiosi, C., Fagotto, F., 1994, ApJS 94, 63 
 
\bibitem[1993]{bruzual} Bruzual, G.A., Charlot, S., 1993, ApJ 405, 538 
 
\bibitem[1995]{buta} Buta, R., Mitra, S., de Vaucouleurs, G., Corwin, H.G. Jr., 1995, AJ 107, 118 
 
\bibitem[Carigi, Col{\'{\i}}n, \& Peimbert(1999)]{1999ApJ...514..787C} 
Carigi, L., Col{\'{\i}}n, P., \& Peimbert, M.\ 1999, \apj, 514, 787 

\bibitem[1994] {carollo} Carollo, C. M., Danziger, I. J., 1994, MNRAS 270, 523 $+$ 743 
 
\bibitem[1997]{chabrier} Chabrier, G., Baraffe, I., 1997, A\&A 327, 1039 
 
\bibitem[1997]{coleman} Coleman, G.D., Wu, C.C., Weedman, D.W., 1980, ApJS 43, 393 
 
\bibitem[1995]{} Einsel, C., Fritze -- v. Alvensleben, U., Kr\"uger, H., Fricke, K. J., 1985, A\&A 296, 347 
 
\bibitem[1998] {a} Ferguson, A. M. N., Gallagher, J. S., Wyse, R. F. G., 1998, AJ 116, 673  
 
\bibitem[2000]{ferguson} Ferguson, H.C., Dickinson, M., Williams, R., 2000, ARA\&A, 38, 667
 
\bibitem[1997]{fioc} Fioc, M., Rocca -- Volmerange, B., 1997, A\&A 326, 950 
 
\bibitem[1999]{fioc9} Fioc, M., Rocca -- Volmerange, B., 1999, A\&A 351, 869 
 
\bibitem[1994]{fritze} Fritze -- v. Alvensleben, U., Gerhard, O.E., 1997, A\&A 285, 751 
 
\bibitem[1999]{gardner} Gardner, J.P., et al., 2000, AJ, 119, 486
 
\bibitem[1998]{b} Guiderdoni, B., Rocca -- Volmerange, B., 1988, A\&AS 74, 185 
 
\bibitem[1997]{c} Guiderdoni, B., Rocca -- Volmerange, B., 1987, A\&A 186, 1 
 
\bibitem[1997]{vandenhoek} van den Hoek, L.B., Groenewegen, M.A.T., 1997, A\&AS 123, 305 
 
\bibitem[1997]{hu} Hu, E.M., McMahon, R.G., Cowie, L.L., 1999, ApJ 522, L9 
 
 
\bibitem[1992]{kennicutt} Kennicutt (jr.), R. C., 1992, ApJS 79, 255 
 
\bibitem[1998]{kennicutt8} Kennicutt (jr.), R. C., 1998, AR A\&A 36, 189 
 
\bibitem[1994] {d} Kilian -- Montenbruck, J., Gehren, T., Nissen, P. E., 1994, A\&A 291, 757 
 
\bibitem[Kroupa, Tout, \& Gilmore(1993)]{1993MNRAS.262..545K} Kroupa, P., 
Tout, C.~A., \& Gilmore, G.\ 1993, \mnras, 262, 545

\bibitem[1992]{kurucz} Kurucz, R.L., 1992, in ``The Stellar Populations  
        of Galaxies'', IAU Symp. \#~149, eds. B. Barbury, A. Renzini,  
        Kluwer, Dordrecht, p.225 
         
\bibitem[1982] {lamla} Lamla, E., 1982, in Landolt-B\"ornstein, eds. K.-H. Hellwege, Springer, 
        Band VI/2b, p. 73 
 
\bibitem[1997] {lejeune7} Lejeune, T.,  Cuisinier, F., Buser, R., 1997, A\&AS 125, 229 
 
\bibitem[1998] {lejeune8} Lejeune, T.,  Cuisinier, F., Buser, R., 1998, A\&AS 130, 65 
 
\bibitem[1999]{lindner9} Lindner, U., Fritze -- v. Alvensleben, U., Fricke, K. J., 1999, A\&A 341, 709 
 
\bibitem[1996]{lindner6} Lindner, U., Fritze -- v. Alvensleben, U., Fricke, K. J., 1996, A\&A 316, 123 
 
\bibitem[1999]{loewenstein} Loewenstein, M., 1999, in {\sl Star Formation in Early-Type 
        Galaxies}, eds P. Carral, J. Cepa, ASP Conf. Ser., {\sl in press} 
 
\bibitem[1997]{loewenthal} Lowenthal, J. D., et al., 1997, ApJ 481, 673 
 
\bibitem[\protect\citeauthoryear{Madau}{1995}]{madau} Madau, P., 1995, ApJ 441, 18 
 
\bibitem[1994] {f} McWilliam, A., Rich, R. M., 1994, ApJS 91, 749 

\bibitem[M{\" o}ller, Fritze-v.~Alvensleben, Calzetti, \& 
Fricke(2001)]{2001IAUS..204..413M} M{\" o}ller, C.~S., 
Fritze-v.~Alvensleben, U., Calzetti, D., \& Fricke, K.~J.\ 2001a, IAU 
Symposium, 204, 413 

\bibitem[M{\" o}ller, Fritze-v.~Alvensleben, \& 
Calzetti(2001)]{2001ApSSS.277..601M} M{\" o}ller, C.~S., 
Fritze-v.~Alvensleben, U., \& Calzetti, D.\ 2001b, Astrophysics and Space 
Science Supplement, 277, 601 

\bibitem[M{\" o}ller, Fritze-V.~Alvensleben, Fricke, \& 
Calzetti(2001)]{2001Ap&SS.276..799M} M{\" o}ller, C.~S., 
Fritze-V.~Alvensleben, U., Fricke, K.~J., \& Calzetti, D.\ 2001c, \apss, 
276, 799 

\bibitem[1996] {g} M\"oller, C. S., Fritze -- v. Alvensleben, U., Fricke, K. J., 1996, In: From Stars 
        to Galaxies, eds. C. Leitherer et al., ASP Conf. Ser. 98, p. 496   
 
\bibitem[1997] {h} M\"oller, C. S., Fritze -- v. Alvensleben, U., Fricke, K. J., 1997, A\&A 317, 676  
 
\bibitem[1999] {i} M\"oller, C. S., Fritze - v. Alvensleben, U., Fricke, K. J., 1999a,  
        in {\sl The Birth of Galaxies}, {\sl in press}  
 
\bibitem[1999] {j} M\"oller, C. S., Fritze - v. Alvensleben, U., Fricke, K. J., Calzetti, D., 1999b,  
        in {\sl The Evolution of Galaxies on Cosmological Timescales}, ed. J. Beckman, astro-ph/9906328 
 
\bibitem[1993] {k} Oey, M. S., Kennicutt, R. C. jr., 1993, ApJ 411, 137 
 
\bibitem[2000] {pettini0} Pettini, M., Steidel, C.C., Adelberger, K.L., Dickinson, M., Giavalisco, M., 
        2000, ApJ 528, 96 
 
\bibitem[1999] {pettini} Pettini, M., Ellison, S. L., Steidel, C. C., Bowen, D. V., 1999, ApJ, 510, 576 
 
\bibitem[1997] {l} Pettini, M., King, D.L., Smith, L.J., Hunstead, R.W., 1997, ApJ, 510, 576 
 
\bibitem[1996] {m} Phillipps, S., Edmunds, M. G., 1996, MN 281, 362 
 
\bibitem[1997]{poggianti} Poggianti, B. M., 1997, A\&AS 122, 399 
 
%\bibitem[1998]{pozzetti} Pozzetti, L., Madau, P., Zamorani, G., Ferguson, H.C., Bruzual, G.A., 1998, MNRAS 298, 1133 
 
\bibitem[2000]{ramirez} Ramirez, C., Solange V., Stephens, Andrew W., Frogel, Jay A., DePoy, 
        D. L., 2000 AJ 120, 633 
 
\bibitem[1995]{richer} Richer, M. G., McCall, M., 1995, ApJ 445, 642  
 
\bibitem[1998]{rocha} Rocha - Pinto, H. J., Maciel, W. J., 1998, A\&A 339, 791 
 
\bibitem[1996]{sadler} Sadler, E. M., Rich, R. M., Terndrup, D. M., 1996, AJ 112, 171 
 
\bibitem[1985]{sandagea} Sandage, A., Binggeli, B., Tamman, G.A., 1985a, AJ 90, 395 
 
\bibitem[1985]{sandageb} Sandage, A., Binggeli, B., Tamman, G.A., 1985b, AJ 90, 1759 
 
\bibitem[1997]{sawicki} Sawicki, M., Yee, H.K.C., 1997, AJ 113, 1 

\bibitem[Scalo(1986)]{1986FCPh...11....1S} Scalo, J.~M.\ 1986, Fundamentals 
of Cosmic Physics, 11, 1 
 
\bibitem[Schulz, Fritze-v.~Alvensleben, M{\" o}ller, \& 
Fricke(2002)]{2002A&A...392....1S} Schulz, J., Fritze-v.~Alvensleben, U., 
M{\" o}ller, C.~S., \& Fricke, K.~J.\ 2002, \aap, 392, 1 

 
\bibitem[2000]{teplitz} Teplitz, H.I., et al., 2000, ApJ, 542, 18 
 
\bibitem[1997]{trager} Trager, S. C., Faber, S. M., Dressler, A., Oemler, A., 1997, ApJ 485, 92 
 
 
\bibitem[1996]{williams} Williams, R.E., et al., 1996, AJ 112, 1335 
 
\bibitem[1996]{williams1} Williams, R.E., et al., 1998, Am. Astron. Society Meeting 193, 7501 
 
\bibitem[Vazdekis, Peletier, Beckman, \& Casuso(1997)]{1997ApJS..111..203V} 
Vazdekis, A., Peletier, R.~F., Beckman, J.~E., \& Casuso, E.\ 1997, \apjs, 
111, 203 

\bibitem[Vazdekis, Casuso, Peletier, \& Beckman(1996)]{1996ApJS..106..307V} 
Vazdekis, A., Casuso, E., Peletier, R.~F., \& Beckman, J.~E.\ 1996, \apjs, 
106, 307 

\bibitem[V{\' a}zquez, Carigi, \& Gonz{\' a}lez(2003)]{2003A&A...400...31V} 
V{\' a}zquez, G.~A., Carigi, L., \& Gonz{\' a}lez, J.~J.\ 2003, \aap, 400, 
31
         
\bibitem[1994]{zaritsky} Zaritsky, D., Kennicutt, R. C., Huchra, J. P., 1994, ApJ 420, 87 
 
\bibitem[1998]{vanzee} van Zee, L., Salzer, J. J., Haynes, M. P., \etal, 1998, AJ 116, 2805 
 
\end{thebibliography}
\end{document}